\documentclass[%
reprint,
superscriptaddress,
groupedaddress,
nobibnotes,
 amsmath,amssymb,
 aps,
prb,
]{revtex4-2}
\usepackage[dvipsnames]{xcolor}
\usepackage{graphicx}% Include figure files
\usepackage{dcolumn}% Align table columns on decimal point
\usepackage{bm}% bold math
\usepackage[breaklinks=true,colorlinks,citecolor=blue,linkcolor=red,urlcolor=blue]{hyperref}
\usepackage{braket}
\usepackage[utf8]{inputenc} % special hats
\usepackage{mathtools}
\usepackage[bottom]{footmisc}
\usepackage{hyperref}
\usepackage{amsfonts}
\usepackage{amsmath}
\usepackage{slashed}
\usepackage{amsthm}
\usepackage{mathbbol}
\usepackage{soul}

\usepackage{tikz}

\usepackage{tikz}

\makeatletter
\let\orig@addcontentsline\addcontentsline

\newcommand{\enableSMtoc}{%
  \renewcommand{\addcontentsline}[3]{%
    \orig@addcontentsline{##1}{##2}{##3}%
    \def\tempa{##1}\def\tempb{toc}%
    \ifx\tempa\tempb
      \orig@addcontentsline{smtoc}{##2}{##3}%
    \fi
  }%
}

\newcommand{\SMtableofcontents}{\@starttoc{smtoc}}
\makeatother
\begin{document}

%\preprint{APS/123-QED}

\title{Helical Domain-Wall-Ring Networks Reshape Superconducting Correlations}

%% Authors %%%
\author{Yi-Chun Hung}
\email[Contact author: ]{hung.yi@northeastern.edu}

\author{Arun Bansil}
\email[Contact author: ]{ar.bansil@northeastern.edu}

\affiliation{Department of Physics,\;Northeastern\;University,\;Boston,\;Massachusetts\;02115,\;USA}
\affiliation{Quantum Materials and Sensing Institute,\;Northeastern University,\;Burlington,\;Massachusetts\;01803,\;USA}


\begin{abstract}
Extended domain-wall networks that emerge in moiré materials provide a distinct platform for quasi-one-dimensional electronic states. However, the interaction driven orders in confined networks remain largely unexplored. Here, we discuss superconducting (SC) correlations in interacting helical domain-wall-ring networks in the closed topological domains formed within the moiré patterns of an underlying twisted bilayer honeycomb lattice. We first analyze the system within the framework of an infinite-size theory and show that inter-ring SC-pair tunneling is renormalization-group relevant {in an appropriate interaction regime, favoring} enhanced SC correlations through inter-ring phase locking. To address finite-size effects resulting from the ring-network geometry, we present a self-consistent variational approach. Our analysis shows that even in the regime where the infinite-size theory predicts strongly coupled pair tunneling, the induced phase-locking scale remains strongly suppressed. In contrast, the SC scaling dimension {is governed mainly by density-density interactions and ring geometry, decreasing modestly with decreasing twist angle} while remaining insensitive to the pair-tunneling strength. This discrepancy demonstrates that ring networks {cannot generally be understood as simple finite-size truncations of} their infinite-size counterparts but can exhibit qualitatively distinct collective behavior. Our study highlights how the interplay of confinement effects and ring-network geometry can reshape SC correlations.
\end{abstract}

\maketitle
% ==============================================
\section{Introduction}
\par Moiré materials provide a versatile platform for engineering electronic structures and interaction-driven phases. In addition to their nearly flat bands that host exotic phenomena such as unconventional superconducting (SC) orders \cite{Balents2020, PhysRevX.8.041041, doi:10.1126/science.aav1910, Cao2018, Cao2018_2, Oh2021, Guo2025} and fractional quantum anomalous Hall states \cite{PhysRevX.1.021014, PhysRevX.13.031037, PhysRevB.108.085117, Park2023, Cai2023}, the moiré pattern at small twist angles can generate one-dimensional (1D) states that form extended networks. Prominent examples include the domain-wall network in twisted bilayer graphene (TBG) arising from the quantum valley Hall effect \cite{PhysRevB.99.161405, PhysRevB.101.165431, PhysRevResearch.1.013001, doi:10.1073/pnas.1309394110, Wang_2024_TBG_net, Gargiulo_2018}, the quasi-1D low-energy states in twisted-bilayer black phosphorus \cite{PhysRevB.108.L201120, PhysRevB.96.195406, doi:10.1073/pnas.2527371123, D1NR07736H, doi:10.1126/science.adq2977, Cao2016}, and the M-valley transition-metal dichalcogenides \cite{PhysRevLett.135.196402, calugaru2024MTMD, jiang20242database}. However, these systems are typically modeled as arrays of coupled 1D channels of infinite length. The role of confinement and network geometry on the interaction-driven orders remains largely unexplored.

\par The low-energy physics of interacting 1D conducting electrons is generally described by the Tomonaga-Luttinger-liquid theory \cite{Tomonaga_1950, Luttinger_1963, LL_correction_1965}, which applies to systems ranging from quantum wires to topological edge states. In particular, the helical edge states of quantum spin Hall effect (QSHE) realize a helical Luttinger liquid (HLL) \cite{HLL_review_2021, HLL_review_2026}. In the presence of time-reversal symmetry (TRS), the QSHE is characterized by a nontrivial $\mathbb{Z}_2$ invariant \cite{PhysRevLett.95.226801, PhysRevLett.95.146802, PhysRevB.95.075146, PhysRevB.74.195312}, which protects the HLL against elastic single-particle backscattering \cite{PhysRevLett.96.106401}. Owing to the underlying spin-momentum locking, HLLs provide a fertile platform for exotic excitations, such as the Majorana bound states \cite{PhysRevB.102.205402, PhysRevB.100.165420, PhysRevB.101.104502, PhysRevLett.121.186801, PhysRevLett.121.096803, PhysRevB.106.085420, D4NH00254G} and parafermion excitations \cite{PhysRevB.90.155447, PhysRevLett.113.036401, PhysRevB.91.081406, PhysRevLett.122.066801, PhysRevB.110.155117}. Their multichannel counterparts support a broad range of emergent phenomena, including Majorana Kramers pairs \cite{PhysRevLett.121.196801, PhysRevB.90.155447, PhysRevB.102.195401, PhysRevB.111.245145} and higher-order topological insulators \cite{PhysRevB.110.035125, PhysRevB.108.245103, alpha_Sb}. Despite extensive studies of isolated HLLs and a few coupled-HLL systems \cite{PhysRevB.90.035116, PhysRevB.95.045150, PhysRevB.99.045125, Phys.Rev.B.113.075125, PhysRevB.112.19527}, the physics of moiré-induced HLL networks largely remains an open question.

\par {Moiré superlattices can host closed topological domains surrounded by trivial background, with helical states along their boundaries giving rise to \emph{helical domain-wall rings}, as exemplified by twisted bilayer Bi$_2$(Te$_{1-x}$Se$_x$)$_3$ \cite{PhysRevResearch.4.043045} and a related effective model \cite{PhysRevB.107.125308}. More recently, structural relaxation in twisted bilayer BiSb has been shown to stabilize and enlarge such topological domains as the twist angle decreases \cite{bordoloi2026}.} This ring network thus provides a natural platform for exploring the interplay of confinement, network geometry, and interactions.

\par In this work, we investigate the collective many-body physics of interacting helical domain-wall rings in a twisted bilayer honeycomb lattice. In the small-twist-angle limit, where the adjacent domain-wall segments become sufficiently long, we approximate the system as coupled infinite-length HLLs and employ a renormalization-group (RG) analysis to examine the relevance of inter-ring pair tunneling and its impact on the SC scaling dimension. {To explicitly account for the closed-ring boundary conditions and network geometry at finite twist angles,} we develop a finite-size theory of HLL on domain-wall rings and employ a self-consistent variational approach to examine whether the RG-predicted SC correlation persists in the finite-size ring network.

% ==============================================
\section{Helical domain-wall rings} 
\begin{figure}[t]
  \centering
  \centering
    \includegraphics[width=\linewidth]{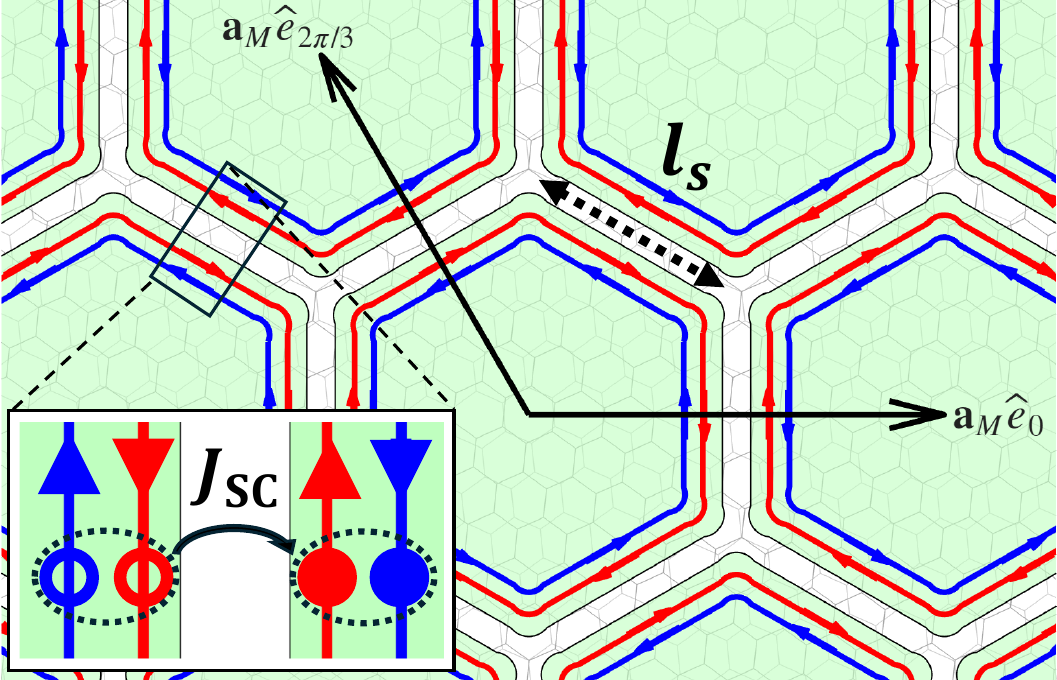}
  \caption{A schematic of the helical domain-wall-ring network arising from closed topological domains (green regions) within the moiré pattern, embedded in a trivial background (white regions). Red (blue) arrows denote counterclockwise (clockwise) helical domain-wall states. Black arrows indicate the moiré lattice vectors of length $a_M$, while the dashed line marks the hexagon side length $l_s$. $\hat{e}_{\theta}\equiv \cos(\theta)\hat{e}_x+\sin(\theta)\hat{e}_y$. Inset: SC-pair tunneling with strength $J_{\rm SC}$ between the neighboring domain walls; empty (filled) circles represent holes (particles).
  }
  \label{fig:01}
\end{figure}

We begin by modeling the HLLs residing on closed helical domain-wall rings, which arise from closed topological domains formed within the moiré pattern of a twisted bilayer honeycomb lattice, with their size determined by the moiré period. At small twist angles, the enlarged domains are expected to form a nearly space-filling pattern, motivating an effective description in which each helical domain-wall ring can be approximated as a hexagon with rounded corners (Fig.~\ref{fig:01}). The hexagon consists of six nearly straight segments of length $l_s\approx L/6\approx a_M/\sqrt{3}$ \footnote{In the small twist-angle limit, the size of the rounded corner $l_c$ remains much smaller than the length of the connecting domain-wall segment, $l_c \ll L$. The corner region, therefore, contributes only local corrections and can be neglected in the low-energy physics.}, where $L$ and $a_M$ denote the ring circumference and moiré superlattice period, respectively. We model the helical domain-wall states as,
\begin{equation}\label{eq:left-right-modes}
\psi_{\mathbf{R}}(r)=\mathcal{U}_{\mathbf{R},+,\uparrow}(r)e^{ik_F r}
+\mathcal{U}_{\mathbf{R},-,\downarrow}(r)e^{-ik_F r},
\end{equation}
where $r\in[0,L)$ denotes the counterclockwise coordinate along the domain wall, $k_F$ the Fermi momentum, and $\mathbf{R}$ the center of the domain-wall ring.  Owing to the ring geometry, $\psi_{\mathbf{R}}(r+L)=\psi_{\mathbf{R}}(r)$. The slow varying fields $\mathcal{U}_{\mathbf{R},+,\uparrow}(r)$ and $\mathcal{U}_{\mathbf{R},-,\downarrow}(r)$ correspond to clockwise- and counterclockwise-moving modes, respectively, whose spins are locked to propagation directions. {This spin assignment is merely a convention adopted to specify a configuration for calculations without loss of generality.} We omit the spin index for simplicity hereafter.

\par The combination of spin-momentum locking, TRS, and the ring-network geometry strongly constrains the allowed inter-ring interactions. We will show in the following section that SC-pair tunneling is the leading symmetry-allowed process here. Therefore, we focus on SC correlations in helical domain-wall-ring networks. For this purpose, we first model each ring as a finite-length HLL described by the Hamiltonian,
\begin{equation}\label{eq:H_tot}
    H_{\rm HLL} = \sum_{\mathbf{R}} H_{0,\mathbf{R}} + H_{\text{intra},\mathbf{R}}.
\end{equation}
Here, $H_{0,\mathbf{R}}$ denotes the single-particle Hamiltonian:
\begin{equation}
H_{0,\mathbf{R}} = -i\hbar v_F\int_0^L dr  \sum_{\xi=\pm}\xi \mathcal{U}_{\mathbf{R},\xi}^{\dagger}(r)\partial_r\mathcal{U}_{\mathbf{R},\xi}(r).
\end{equation}
$H_{\text{intra},\mathbf{R}}$ denotes the intra-ring density-density interactions:
\begin{align}
H_{\text{intra},\mathbf{R}} = & \int_0^L\int_0^L drdr' V(r-r')\rho_{\mathbf{R}}(r)\rho_{\mathbf{R}}(r'),
\end{align}
where $v_F$, $\rho_{\mathbf{R}}(r)=\sum_{\xi=\pm}\mathcal{U}_{\mathbf{R},\xi}^{\dagger}(r)\mathcal{U}_{\mathbf{R},\xi}(r)$, and $V(r)$ denote the Fermi velocity, charge density, and the short-ranged intra-ring interaction potential. Bosonizing Eq.~\eqref{eq:H_tot} {via a standard procedure \cite{book:Giamarchi}} yields
\begin{align}
    H_{\rm HLL} = \sum_{\mathbf{R}} \frac{\hbar u}{2\pi}\int_0^L dr \frac{1}{K}(\partial_r\varphi_{\mathbf{R}}(r))^2 + K(\partial_r\vartheta_{\mathbf{R}}(r))^2,
\end{align}
where $u = v_F\left(1+\frac{2V_0}{\pi\hbar v_F}\right)^{1/2}$ and the HLL parameter is
\begin{equation}
K = \left( 1+\frac{2V_0}{\pi\hbar v_F} \right)^{-\frac{1}{2}}.
\end{equation}
Here, $V_0$ is the zero-momentum component of $V(r)${, retained under the standard long-wavelength approximation for sufficiently short-ranged interactions \cite{book:Giamarchi}. The bosonic fields $\varphi_{\mathbf R}(r)$ and $\vartheta_{\mathbf R}(r)$ connect to the fermionic fields $\mathcal{U}_{\mathbf{R},\pm}(r)$ via:
\begin{equation}\label{eq:bosonization}
\mathcal{U}_{\mathbf{R},\pm}(r) = \frac{\kappa_\pm}{\sqrt{2\pi a}}e^{i[\vartheta_{\mathbf{R}}(r)\pm\varphi_{\mathbf{R}}(r)]},
\end{equation}
where $a$ is the UV cutoff and $\kappa_\pm$ is the Klein factor satisfying $\kappa_-\kappa_+ = - \kappa_+\kappa_-$ \cite{PhysRevLett.121.196801}}; see {Appendix \ref{sec:appx_a}} for details.

% ==============================================
\section{Inter-ring interactions} We turn now to consider inter-ring interactions, starting with the inter-ring density-density interaction:
\begin{align}
H_{\text{inter}} = & \sum_{\mathbf{R}}\sum_{j=0}^{5}\int_{jl_s}^{(j+1)l_s}\int_{jl_s}^{(j+1)l_s}drdr' U(r-r') \notag
\\ & \quad \quad \quad \quad \times \left[ \rho_{\mathbf{R}}(r){\rho_{\mathbf{R}+\mathbf{a}_j}(f_j(r'))} \right],
\end{align}
where $\mathbf{a}_j = a_M\left[\cos(\frac{j\pi}{3})\mathbf{\hat{e}_x} + \sin(\frac{j\pi}{3})\mathbf{\hat{e}_y}\right]$, $\mathbf{\hat{e}_i}$ is the unit vector along the $i$th direction, and $U(r)$ is the short-ranged inter-ring interaction potential. {With the same counterclockwise coordinate convention on every ring, the shared segment $r\in[jl_s,(j+1)l_s]$ maps to the opposite segment on the neighboring ring according to $f_j(r)=(2j+4)l_s-r\pmod L$. We retain only nearest-neighbor inter-ring interactions, assuming sufficiently strong screening of the Coulomb potential such that interactions between distant rings can be neglected.}

\par In addition to the density-density interactions, domain-wall states on the neighboring rings can also interact through two-particle inter-ring tunneling processes. Since the two neighboring domain walls form an orthohelical configuration {(see inset in Fig.~\ref{fig:01})} \cite{HLL_review_2021}, TRS allows single-particle tunneling only with a rapidly oscillating phase, which makes it irrelevant to the low-energy theory. Nevertheless, a second-order effect can generate a relevant two-particle tunneling term. Due to TRS, only the SC channel admits a symmetry-allowed inter-ring tunneling; see {Appendix \ref{sec:appx_b}} for details. The resulting SC-pair tunneling (inset of Fig.~\ref{fig:01}) is described by:
\begin{align}
H_{\rm T} =  -J_{\rm SC}\sum_{\mathbf{R}}\sum_{j=0}^{5} & \int_{jl_s}^{(j+1)l_s} dr \,\, \mathcal{U}_{\mathbf{R}+\mathbf{a}_j,+}^\dagger(r')\mathcal{U}_{\mathbf{R}+\mathbf{a}_j,-}^\dagger(r') \notag
\\ & \times \mathcal{U}_{\mathbf{R},-}(r)\mathcal{U}_{\mathbf{R},+}(r) + \text{H.c.},
\end{align}
where {$r'=f_j(r)$} and $J_{\rm SC}>0$. The SC-pair tunneling tends to establish SC phase locking between the neighboring domain walls, and thus it reduces the SC scaling dimension in the strong-coupling limit, as detailed below.

\par To focus on the physics introduced by inter-ring interactions, we first consider only the two domain walls on the neighboring domain-wall rings and take $l_s\to\infty$. The resulting {bosonized} Hamiltonian is
\begin{align}
    H = & \sum_{\mu=\pm}\frac{\hbar\tilde{u}_\mu}{2\pi}\int_0^{l_s\to\infty} dr \frac{1}{\tilde{K}_\mu}(\partial_r\varphi_{\mu}^{(\infty)}(r))^2 + \tilde{K}_\mu(\partial_r\vartheta_{\mu}^{(\infty)}(r))^2 \notag
    \\ & - \frac{J_{\rm SC}}{(2\pi a)^2}\int_0^{l_s\to\infty} dr \cos[2\sqrt{2}\vartheta_-^{(\infty)}(r)], \label{eq:inf_H}
\end{align}
where $\varphi_{\pm}^{(\infty)}(r)=[\varphi_{\rm ring\,1}(r)\pm\varphi_{\rm ring\,2}(r)]/\sqrt{2}$, with a similar expression for the $\vartheta_\pm^{(\infty)}(r)$ fields. {These bosonic fields connect to the fermionic fields on the corresponding rings via Eq.~\eqref{eq:bosonization}. As for the intra-ring interaction, we retain only the zero-momentum component $U_0$ of the short-ranged inter-ring potential $U(r)$ under the standard long-wavelength approximation.} Renormalized velocities and HLL parameters are {thus} $\tilde{u}_\pm =v_F(1+\frac{2V_0}{\pi\hbar v_F}\pm\frac{2U_0}{\pi\hbar v_F})^{1/2}$ and
\begin{align}\label{eq:tilde_K}
\tilde{K}_\pm & =\left( 1+\frac{2V_0}{\pi\hbar v_F}\pm\frac{2U_0}{\pi\hbar v_F} \right)^{-\frac{1}{2}}.
\end{align}
Following a standard derivation \cite{PhysRevB.97.045415}, the leading-order RG-flow equation is: $d\tilde{J}_{\rm SC}/dl = 2(1 - \tilde{K}_-^{-1})\tilde{J}_{\rm SC}$. Here, $l\equiv\ln(a(l)/a(0))$, {$a(l)$ is the running UV cutoff with $a(0)$ set to the lattice constant}, and $\tilde{J}_{\rm SC} \equiv J_{\rm SC}/ \hbar u$. Since $l_s\approx L/6\approx a_M/\sqrt{3}$ and $a_M$ varies with the twist angle $\theta_t$, this RG flow equation allows us to relate the twist angle to the onset of the strong-coupling regime through
\begin{equation}\label{eq:theta_RG}
    \theta_{\rm RG} \simeq 2\sin^{-1}\left( \frac{[\tilde{J}_{\rm SC}(0)]^{1/\left[2(1-\tilde{K}_{-}^{-1})\right]}}{2\sqrt{3}} \right).
\end{equation}
When $\theta_t<\theta_{\rm RG}$, $\tilde{J}_{\rm SC}$ reaches the strong-coupling regime within a domain-wall segment. For $\theta_t\gtrsim\theta_{\rm RG}$, however, the RG flow is terminated by the finite segment length before strong coupling is reached. In this work, we neglect the detailed commensurability condition and use the continuum expression $a_M\approx a(0)/[2\sin(\theta_t/2)]$, which is valid for twisted bilayer honeycomb lattices \cite{doi:10.1073/pnas.1108174108}. {Specifically, integrating the RG flow equation gives $\tilde{J}_{\rm SC}(l)=\tilde{J}_{\rm SC}(0)e^{2(1-\tilde{K}_{-}^{-1})l}$. The strong-coupling condition $\tilde{J}_{\rm SC}(l^*)=1$, together with $l_s=a_M/\sqrt{3}=a(0)e^{l^*}$ at $\theta_t=\theta_{\rm RG}$, yields Eq.~\eqref{eq:theta_RG}.} In the following, $\theta_{\rm RG}$ serves as the reference scale for comparison with our finite-size calculations.

\par For $\theta_t<\theta_{\rm RG}$, the pair-tunneling term is expected to enhance SC correlations by pinning the $\vartheta_-^{(\infty)}$ field. To characterize this tendency, we examine the scaling dimension $\eta_{\rm SC}$ associated with the operator characterizing SC instability $\mathcal{O}_{\rm SC}(r)=\mathcal{U}_{\mathbf{R},+}(r)\mathcal{U}_{\mathbf{R},-}(r)-\mathcal{U}_{\mathbf{R},-}(r)\mathcal{U}_{\mathbf{R},+}(r)\sim e^{i2\vartheta_{\mathbf{R}}(r)}$, which characterizes the power-law decay of the SC correlation function $\langle \mathcal{O}_{\rm SC}^\dagger(r)\mathcal{O}_{\rm SC}(0)\rangle \sim r^{-2\eta_{\rm SC}}$. To disentangle the effects of different inter-ring interactions, we introduce three characteristic scaling dimensions. In the decoupled-ring limit, the SC scaling dimension is $\eta_{\rm SC}^{(0)} = K^{-1}$. Including only the inter-ring density-density interaction reduces it to $\eta_{\rm SC}^{(U)}=\left( \tilde K_+^{-1} + \tilde K_-^{-1} \right)/2$. When $\theta_t<\theta_{\rm RG}$ and $\vartheta_-^{(\infty)}$ is pinned, the scaling dimension further reduces to $\eta_{\rm SC}^{(\rm pin)}=\tilde K_+^{-1}/2$. For physically relevant interaction parameters where $2(U_0-V_0)/\pi\hbar v_F<1$ (see {Appendix \ref{sec:appx_c}} for details), these quantities satisfy
\begin{equation}\label{eq:eta_hierarchy}
\eta_{\rm SC}^{(0)} \quad  > \quad  \eta_{\rm SC}^{(U)} \quad  > \quad \eta_{\rm SC}^{(\rm pin)},
\end{equation}
reflecting the successive enhancement of SC correlations by inter-ring density-density interactions and pair-tunneling-induced phase locking. These characteristic values serve as useful benchmarks for interpreting the finite-size results below.

% ==============================================
\section{Self-consistent theory for the ring networks} The preceding analysis is based on the infinite-domain-wall approximation and predicts that the pair-tunneling term drives the system toward a strong-coupling regime for $\theta_t<\theta_{\rm RG}$. Whether this tendency survives at larger twist angles remains an open question. To address this issue, we develop a self-consistent variational theory \cite{feynman1972, book:Giamarchi}.

\par We introduce a Gaussian trial action $S_0$ characterized by a variational Green's function $\mathcal{G}(m_J)$ containing a variational mass parameter $m_J$ such that
\begin{equation}
\mathcal{G}^{-1}(m_J)=\mathcal{G}_{\rm full\,HLL}^{-1}+{\frac{m_J}{(2\pi a)^2}\mathcal{M}}.
\end{equation}
Here, $\mathcal{G}_{\rm full\,HLL}$ is the Green's function associated with the full HLL action $S_{\rm full\,HLL}$, which includes the effects resulting from inter-ring density-density interaction into HLL, and $\mathcal{M}$ encodes the structure of pair tunneling, as detailed below. $m_J$ is determined self-consistently by the stationarity condition of the variational free energy,
\begin{equation}\label{eq:station_main}
\frac{\delta }{\delta \mathcal{G}(m_J)}
\left[ F_0+ T\langle S-S_0\rangle_0 \right]=0,
\end{equation}
where $F_0$ and $\langle\cdots\rangle_0$ are the free energy and expectation value associated with the trial action $S_0$, respectively, $S$ is the action of the original system, and $T$ is the temperature. A nonzero $m_J$ thus indicates inter-ring phase locking and sets its characteristic energy scale. 

\par To formulate the self-consistent variational theory, we mode-expand the bosonic fields $\mathcal{F}_{\mathbf{R}}(r,\tau) = \frac{1}{\sqrt{NL}}\int\frac{d\omega}{2\pi}\sum_{\mathbf{p}=(\mathbf{q},k_n)}e^{i\mathbf{p}\cdot\mathbf{\mathcal{R}}-i\omega\tau}\mathcal{F}_{\mathbf{p}}(\omega)$, where $\mathcal{F}_{\mathbf{R}}=\varphi_{\mathbf{R}},\vartheta_{\mathbf{R}}$, $\mathbf{\mathcal{R}}=(\mathbf{R},r)$, and $\mathbf{p}=(\mathbf{q},k_n)$ combines the moiré crystal momentum $\mathbf{q}$ and the ring-harmonic momentum $k_n=2\pi n/L$ ($1\le |n|\lesssim \lfloor L/2a\rfloor$). Here, $N$ is the number of domain-wall rings and $\omega$ denotes the Matsubara frequency in the $T\to0$ limit. Zero modes are excluded, as they do not qualitatively alter our conclusions; see Supplemental Material (SM) for details \cite{SM}\nocite{LL_Haldane_1981, senechal1999}. Then, $S_{\rm full\,HLL}$ can be expressed as
$S_{\rm full\,HLL} = \frac{1}{2}\int \frac{d\omega}{2\pi}\sum_{\mathbf{q}}\boldsymbol{\vartheta}_{\mathbf{q}}^{\dagger}(\omega)\mathcal{G}_{\rm full\,HLL,\mathbf{q}}^{-1}(\omega)\boldsymbol{\vartheta}_{\mathbf{q}}(\omega)$
Here, $[\boldsymbol{\vartheta}_{\mathbf{q}}(\omega)]_{n} = \vartheta_{\mathbf{q},k_n}(\omega)$ and
\begin{align}
\mathcal{G}_{{\rm full\,HLL},\mathbf{q}}^{-1}(\omega) & = \frac{1}{\pi\hbar u}\omega^2\mathcal{K} B_{\mathbf{q}}^{-1}\mathcal{K} + \frac{\hbar u}{\pi}A_{\mathbf{q}},
\end{align}
where $[A_{\mathbf{q}}]_{nm} = K(k_n)^2\delta_{nm}$, $[\mathcal{K}]_{nm}=k_n\delta_{nm}$, and
\begin{align}
[B_{\mathbf{q}}]_{nm} = & {k_nk_m\bigg[ \frac{1}{K}\delta_{nm} + \frac{2U_0}{\pi\hbar u}\frac{\sin(\frac{\pi (n+m)}{6})}{\pi (n+m)}} \notag
\\ & {\times\sum_{j=0}^{5}e^{-i\frac{\pi}{6}(2j+4)(n-m)}\cos\left(\mathbf{q}\cdot\mathbf{a}_j + \frac{\pi}{2}(n+m)\right) \bigg]}. \label{eq:U_q}
\end{align}
The factor $\sin[\pi(n+m)/6]/[\pi(n+m)]$ should be understood as its limiting value $1/6$ when $n+m=0$ (see SM for details \cite{SM}). As indicated by Eq.~\eqref{eq:U_q}, the inter-ring density-density interaction couples different harmonics and incorporates a moiré-scale spatial modulation. These features distinguish the present system from the coupled-HLL theory for infinite-length domain walls.

\par Motivated by the quadratic expansion of the pair-tunneling term around its energy minimum, we model the variational mass term $\mathcal{M}$ as
\begin{align}
    [\mathcal{M}_{\mathbf{q}}]_{nn'} & = \sum_{j=0}^{5}\int_{jl_s}^{(j+1)l_s}\frac{dr}{L}{\Lambda_{j,n'}(\mathbf{q};r)\Lambda_{j,-n}(-\mathbf{q};r)}, \label{eq:trial_G}
\end{align}
where ${\Lambda_{j,n}(\mathbf{q};r) = e^{i\mathbf{q}\cdot\mathbf{a}_j}e^{ik_n f_j(r)}-e^{ik_n r}}$ encodes the spatial structure of the pair-tunneling term in the ring-network geometry. Computing Eq.~\eqref{eq:station_main} by using Eq.~\eqref{eq:trial_G} then yields a self-consistent equation for $m_J$:
\begin{align}
    m_J & = J_{\rm SC}\frac{\sum_{\mathbf{q}}\text{Tr}[\mathcal{M}_{\mathbf{q}}^\dagger\tilde{\mathcal{M}}_{\mathbf{q}}(m_J)]}{\sum_{\mathbf{q}}\text{Tr}[\mathcal{M}_{\mathbf{q}}^\dagger\mathcal{M}_{\mathbf{q}}]}. \label{eq:self_mj}
\end{align}
Here, $\tilde{\mathcal{M}}_{\mathbf{q}}(m_J)$ is an effective tunneling matrix dressed by a fluctuation prefactor $e^{-2C_\Theta(r)}$ {inside the integral of Eq.~\eqref{eq:trial_G}}, where
\begin{equation}
C_\Theta(r) = \left\langle[{\vartheta_{\mathbf{R}+\mathbf{a}_j}(f_j(r))-\vartheta_{\mathbf{R}}(r)}]^2\right\rangle_0,
\end{equation}
See SM for details \cite{SM}.

\begin{figure}[t]
  \centering
  \centering
    \includegraphics[width=\linewidth]{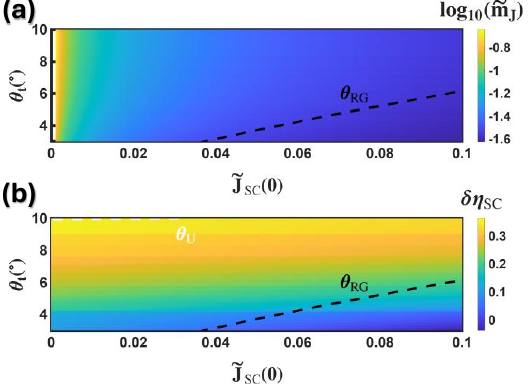}
  \caption{Self-consistently determined (a) tunneling-induced mass $\tilde{m}_J=m_J/m_J^{(\rm pin)}$ and (b) relative deviation of the SC scaling dimension from the pinned-limit value, $\delta\eta_{\rm SC}=\left(\eta_{\rm SC}-\eta_{\rm SC}^{(\rm pin)}\right)/\left(\eta_{\rm SC}^{(0)}-\eta_{\rm SC}^{(\rm pin)}\right)$, as functions of the bare pair-tunneling strength $\tilde{J}_{\rm SC}(0)$ and twist angle $\theta_t$ for $2V_0/\pi\hbar v_F=1.18$ and $2U_0/\pi\hbar v_F=2.08$. Black and white dashed lines denote $\theta_{\rm RG}$ and {$\theta_{\rm U}\approx9.8^\circ$}, respectively.}
  \label{fig:03}
\end{figure}

% ==============================================
\section{Finite-size effects} We next examine the self-consistent results for the induced phase-locking scale $m_J$ and the SC scaling dimension $\eta_{\rm SC}$. $\eta_{\rm SC}$ is obtained from fitting the SC correlation $\langle \mathcal{O}_{\rm SC}^{\dagger}(r)\mathcal{O}_{\rm SC}(0)\rangle_0$ with the finite-size scaling form $\left[L\sin\left(\pi r/L\right)/\pi a\right]^{-2\eta_{\rm SC}}$ \cite{Cardy_1984}. We compare $\eta_{\rm SC}$ and $m_J$ with their infinite-size counterparts $\eta_{\rm SC}^{(\rm pin)}$ and {$m_J^{(\rm pin)} \simeq 2\pi\hbar u\tilde{K}_{-}[\tilde{J}_{\rm SC}(0)]^{2/\left[2(1-\tilde{K}_{-}^{-1})\right]}$} (see {Appendix \ref{sec:appx_d}} for details). The results for $\tilde{m}_J=m_J/m_J^{(\rm pin)}$ and $\delta\eta_{\rm SC}=\left(\eta_{\rm SC}-\eta_{\rm SC}^{(\rm pin)}\right)/\left(\eta_{\rm SC}^{(0)}-\eta_{\rm SC}^{(\rm pin)}\right)$ as functions of $\tilde{J}_{\rm SC}(0)$ and $\theta_t$ are summarized in Figs.~\ref{fig:03}(a) and (b), respectively. We fix $2V_0/\pi\hbar v_F=1.18$ and $2U_0/\pi\hbar v_F=2.08$, for which $\theta_{\rm RG}$ lies within the numerically accessible range. A more comprehensive scan over physically relevant interaction parameters with fixed $\tilde{J}_{\rm SC}(0)$ and $\theta_t$ is presented later in this section.

\par As shown in Fig.~\ref{fig:03}(a), $\tilde{m}_J$ remains finite {for $\tilde{J}_{\rm SC}(0)\neq0$ over the considered parameter regime}, indicating that the pair-tunneling term continues to induce {a finite phase-locking tendency} between the neighboring rings. However, {$\tilde{m}_J$ generally remains below unity and can be appreciably suppressed}, even for $\theta_t<\theta_{\rm RG}$, where the infinite-size theory predicts an RG flow toward strong coupling. Moreover, $\tilde{m}_J$ decreases as $\theta_t$ decreases. 

\par Despite the {reduction of $\tilde{m}_J$}, the extracted SC scaling dimension $\eta_{\rm SC}$ exhibits a markedly different behavior. As shown in Fig.~\ref{fig:03}(b), $\eta_{\rm SC}$ remains {largely} insensitive to $\tilde J_{\rm SC}(0)$ throughout the parameter range considered and decreases monotonically as the twist angle is reduced. {Remarkably, this enhancement of SC correlations occurs even as $\tilde{m}_J$ decreases, indicating that the SC scaling dimension need not directly track the strength of pair-tunneling-induced phase locking, in contrast to the conventional infinite-size strong-coupling picture.} In particular, $\eta_{\rm SC}$ {approaches, and can even fall slightly below, the infinite-size strong-coupling benchmark $\eta_{\rm SC}^{(\rm pin)}$ at small twist angles, despite the reduced $\tilde{m}_J$. Meanwhile, as the twist angle increases, $\eta_{\rm SC}$} undergoes a crossover relative to $\eta_{\rm SC}^{(U)}$. The latter defines a characteristic twist angle $\theta_{\rm U}$ via $\eta_{\rm SC}(\theta_{\rm U})=\eta_{\rm SC}^{(U)}$. The resulting $\theta_{\rm U}$ line separates the regions with $\eta_{\rm SC}>\eta_{\rm SC}^{(U)}$ and $\eta_{\rm SC}<\eta_{\rm SC}^{(U)}$.

\begin{figure}[t]
  \centering
  \centering
    \includegraphics[width=\linewidth]{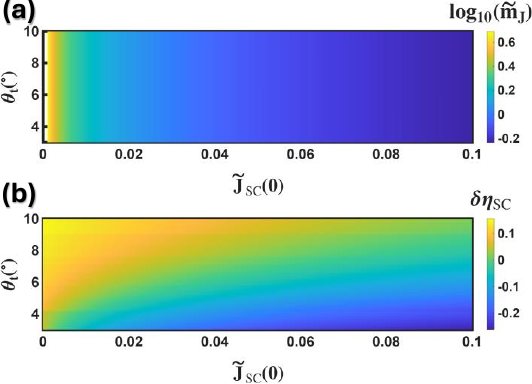}
  \caption{ {Self-consistently determined (a) tunneling-induced mass $\tilde{m}_J=m_J/m_J^{(\rm pin)}$ and (b) relative deviation of the SC scaling dimension from the pinned-limit value, $\delta\eta_{\rm SC}=(\eta_{\rm SC}-\eta_{\rm SC}^{(\rm pin)})/\eta_{\rm SC}^{(\rm pin)}$, as functions of the bare pair-tunneling strength $\tilde{J}_{\rm SC}(0)$ and twist angle $\theta_t$ for $2V_0/\pi\hbar v_F=1.18$ and $2U_0/\pi\hbar v_F=2.08$, obtained by setting $C_{\Theta}(r)=0$ in the self-consistent calculations.} }
  \label{fig:M_no_C}
\end{figure}

\par These trends can be traced to the enhancement of $C_\Theta(r)$, which becomes stronger as the increasing moiré length scale generates more soft modes and the inter-ring interaction further softens them. Consequently, fluctuations associated with the finite-size ring geometry substantially suppress $\tilde{m}_J$ {relative to its infinite-size estimate. To further support this interpretation, we performed an auxiliary self-consistent calculation with $C_\Theta(r)=0$, which brings both $m_J$ and $\eta_{\rm SC}$ close to their respective pinned-limit values, as shown in Fig.~\ref{fig:M_no_C}. This comparison highlights the strong suppression of pair tunneling and its influence on SC correlations by fluctuations.}

\begin{figure}[t]
  \centering
  \centering
    \includegraphics[width=\linewidth]{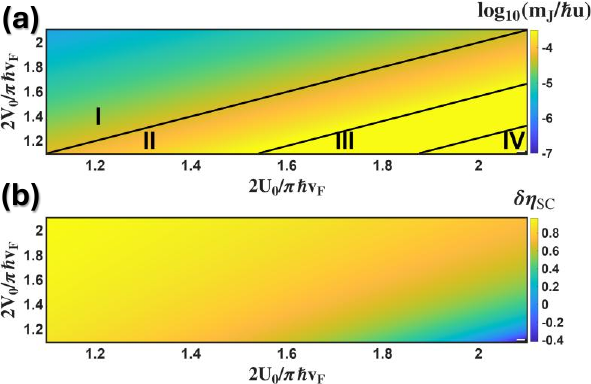}
  \caption{Self-consistently determined (a) $m_J/\hbar u$ and (b) $\delta\eta_{\rm SC}$ as functions of $2V_0/\pi\hbar v_F$ and $2U_0/\pi\hbar v_F$ at fixed $\tilde{J}_{\rm SC}(0)=0.08$ and $\theta_t=3^\circ$. Regions (I--IV) denote: (I) RG-irrelevant $\tilde{J}_{\rm SC}$; (II,III) $\theta_t>\theta_{\rm RG}$ with $m_J>m_J^{(\rm pin)}$ and $m_J<m_J^{(\rm pin)}$, respectively; and (IV) $\theta_t<\theta_{\rm RG}$ with $m_J<m_J^{(\rm pin)}$.}
  \label{fig:int_scan}
\end{figure}

\par We now investigate how the ring-network geometry reshapes the interaction dependence beyond the representative parameters considered above. Focusing on the physically relevant interaction regime where $1.1\leq2V_0/\pi\hbar v_F,\,2U_0/\pi\hbar v_F\leq2.1$ (see {Appendix \ref{sec:appx_c}} for details), Figs.~\ref{fig:int_scan}(a) and (b) show the self-consistently determined $m_J$ and $\delta\eta_{\rm SC}$, as functions of $2V_0/\pi\hbar v_F$ and $2U_0/\pi\hbar v_F$ at fixed $\tilde{J}_{\rm SC}(0)=0.08$ and $\theta_t=3^\circ$. The labeled regions I--IV in Fig.~\ref{fig:int_scan}(a) distinguish different interaction regimes. In the RG-irrelevant regime (I), the infinite-size theory does not support a finite $m_J^{(\rm pin)}$. In contrast, the ring network retains a nonzero $m_J$, accompanied by a decreasing $\eta_{\rm SC}$ as $U_0-V_0$ increases. For larger $U_0-V_0$, $\tilde{J}_{\rm SC}$ becomes RG relevant but does not reach strong coupling due to the finite-size cutoff ($\theta_t>\theta_{\rm RG}$). In this regime, $m_J$ can either exceed (II) or remain below (III) $m_J^{(\rm pin)}$, while $\eta_{\rm SC}$ continues to decrease as $U_0-V_0$ increases {and remains below $\eta_{\rm SC}^{(\rm U)}$ throughout the considered interaction regime (Fig.~\ref{fig:int_scan}(b))}. For sufficiently large $U_0-V_0$ (IV), {the infinite-size RG flow predicts that $\tilde{J}_{\rm SC}$ reaches strong coupling before the finite-size cutoff} ($\theta_t<\theta_{\rm RG}$). Still, $m_J$ remains suppressed such that $m_J<m_J^{(\rm pin)}$. {At large $U_0-V_0$, $\eta_{\rm SC}$ can even fall below the pinned-limit estimate $\eta_{\rm SC}^{(\rm pin)}$ such that $\delta\eta_{\rm SC}<0$, despite the suppressed $m_J$. The limit $2(U_0-V_0)/\pi\hbar v_F\approx1$ corresponds to the divergence of $\tilde{K}_-$.} This limit, however, requires neighboring domain walls to be much closer than the localization length of the domain-wall states, rendering inter-ring tunneling nonperturbative and beyond the validity of our theoretical framework; see {Appendix \ref{sec:appx_c}} for details.

% ==============================================
\section{Discussion} In contrast to the extended domain-wall networks found in systems such as TBG, often treated as arrays of coupled 1D channels with infinite length, helical domain-wall-ring networks consist of closed loops whose circumference is set by the moiré length scale. Remarkably, {our results show that the SC scaling dimension remains largely insensitive to inter-ring pair tunneling, even when the induced phase-locking scale is strongly suppressed.} Unlike the infinite-domain-wall limit, where interactions only renormalize the symmetric and antisymmetric channels at each momentum, coupling between the neighboring rings mixes different ring harmonics and introduces moiré-scale spatial modulation of the collective modes. This distinct behavior suggests a geometry-driven mechanism for reshaping the connection between the phase-locking scale $m_J$ and the SC scaling dimension $\eta_{\rm SC}$. Unraveling the experimental consequences of the coexistence of a strongly suppressed phase-locking scale and enhanced SC correlations would be an interesting direction for future study.

\par More broadly, helical domain-wall-ring networks provide a unique setting in which confinement, network geometry, and interactions become intertwined at low energies. Our results highlight that finite-size quasi-1D networks {may exhibit substantial deviations from their infinite-size counterparts due to geometry-dependent fluctuations}, opening new directions for exploring collective phenomena in engineered moiré systems featuring finite-size quasi-1D networks.

% ==============================================
\section{Acknowledgment}We are grateful to Chen-Hsuan Hsu for important discussions in connection with the variational method. This work was supported by the National Science Foundation through the Expand-QISE award NSF-OMA-2329067 and benefited from the resources of Northeastern University’s Advanced Scientific Computation Center, the Explorer Cluster, and the Massachusetts Technology Collaborative award MTC-22032.

% ==============================================
\section{Data availability}The data that support the findings of this article are not publicly available upon publication because it is not technically feasible and/or the cost of preparing, depositing, and hosting the data would be prohibitive within the terms of this research project. The data are available from the authors upon reasonable request.

% ==============================================
{\section{Use of generative AI tools}ChatGPT (OpenAI; GPT-6) was used to assist with mathematical consistency checks, revising and debugging code, and manuscript revisions; all results were, however, independently verified by the authors.}

% ==============================================
\begin{appendix}
% ==============================================
\section{Bosonization conventions}\label{sec:appx_a}
\par {We discuss detailed properties of the bosonic fields that appear during bosonization. The bosonic fields defined in Eq.~\eqref{eq:bosonization} satisfy the canonical equal-time commutation relation \cite{book:Giamarchi} 
\begin{equation}
[\varphi_{\mathbf{R}}(r),\vartheta_{\mathbf{R}'}(r')] = -i\frac{\pi}{2}\mathrm{sgn}(r-r')\delta_{\mathbf{R},\mathbf{R}'}.
\end{equation}
This relation ensures fermionic anticommutation within each mode, while the Klein factors defined in Eq.~\eqref{eq:bosonization} enforce anticommutation between different modes and rings.} Correspondingly, the bosonic representation of the charge density is
\begin{equation}
    \rho_{\mathbf{R}}(r) = -\frac{1}{\pi}\partial_r\varphi_{\mathbf{R}}(r).
\end{equation}
% ==============================================
\section{Single-particle inter-ring tunneling}\label{sec:appx_b}
\par We demonstrate how the two-particle inter-ring tunneling can emerge from the second-order effect resulting from the single-particle inter-ring tunneling, although the latter is suppressed in the low-energy theory. Due to time-reversal symmetry, the allowed single-particle inter-ring tunneling is:
\begin{align}
H_{\rm T_s} = & t_\perp\sum_{\xi=\pm}\sum_{\mathbf{R}}\sum_{j=0}^{5} \int_{jl_s}^{(j+1)l_s} dr \,\, e^{i\xi k_F[r-{f_j(r)}]} \notag
\\ & \times \mathcal{U}_{\mathbf{R}+\mathbf{a}_j,\xi}^\dagger({f_j(r)})\mathcal{U}_{\mathbf{R},\xi}(r) + \text{H.c.}.
\end{align}
{Here, $f_j(r)=(2j+4)l_s-r\pmod L$.} Although the single-particle tunneling is irrelevant to the low-energy physics because of its rapidly oscillating phase, its second-order contribution can generate a non-oscillating pair-tunneling term. To see this, we consider the action
\begin{align}
S_{T_s} = & t_\perp \sum_{\xi=\pm}\sum_{\mathbf{R}}\sum_{j=0}^{5}\int d\tau\int_{jl_s}^{(j+1)l_s} dr\,e^{i\xi k_F[r-{f_j(r)}]} \notag
\\ & \times \mathcal{U}_{\mathbf{R}+\mathbf{a}_j,\xi}^{\dagger}({f_j(r)},\tau)\mathcal{U}_{\mathbf{R},\xi}(r,\tau) + \mathrm{H.c.}.
\end{align}
The effective low-energy action is obtained by integrating out the high-energy degrees of freedom perturbatively. Expanding the partition function in powers of $S_{T_s}$ yields
\begin{equation}
S_{\rm eff} = S_0 + \langle S_{T_s}\rangle_0 - \frac{1}{2}\left( \langle S_{T_s}^2\rangle_0 - \langle S_{T_s}\rangle_0^2 \right) + \cdots,
\end{equation}
where $\langle\cdots\rangle_0$ denotes averaging with respect to the unperturbed action $S_0$. Since the first-order contribution contains the rapidly oscillating factor $e^{i2\xi k_F r}$, it averages to zero at long wavelengths. The leading contribution, therefore, arises from the second-order cumulant
\begin{equation}
\delta S_{\rm eff}^{(2)} = -\frac{1}{2}\langle S_{T_s}^2\rangle_0.
\end{equation}
Among the resulting four-fermion terms, the product of the $\xi=+$ and $\xi=-$ tunneling processes contains a non-oscillating contribution,
\begin{widetext}
\begin{align}
\delta S_{\rm eff}^{(2)} \sim & -\frac{t_\perp^2}{2}\sum_{\mathbf{R}}\sum_{j=0}^{5}\int d\tau d\tau'\int dr dr'\,e^{i2k_F(r-r')} {\mathcal{U}_{\mathbf{R}+\mathbf{a}_j,+}^{\dagger}(f_j(r),\tau)}\mathcal{U}_{\mathbf{R},+}(r,\tau){\mathcal{U}_{\mathbf{R}+\mathbf{a}_j,-}^{\dagger}(f_j(r'),\tau')}\mathcal{U}_{\mathbf{R},-}(r',\tau') + \mathrm{H.c.},
\end{align}
\end{widetext}
where we keep terms that involve only two domain-wall rings. In the limit of $r'\simeq r$ and $\tau'\simeq\tau$ such that the oscillating phases cancel, the resulting contribution can be written as
\begin{widetext}
\begin{equation}
\delta S_{\rm eff}^{(2)} = -J_{\rm SC}\sum_{\mathbf{R}}\sum_{j=0}^{5}\int d\tau\int_{jl_s}^{(j+1)l_s} dr\,{\mathcal{U}_{\mathbf{R}+\mathbf{a}_j,+}^{\dagger}(f_j(r),\tau)\mathcal{U}_{\mathbf{R}+\mathbf{a}_j,-}^{\dagger}(f_j(r),\tau)}\mathcal{U}_{\mathbf{R},-}(r,\tau)\mathcal{U}_{\mathbf{R},+}(r,\tau) + \mathrm{H.c.},
\end{equation}
\end{widetext}
where $J_{\rm SC}\propto t_\perp^2$. Therefore, although the single-particle tunneling is suppressed in the low-energy theory, its second-order contribution generates an inter-ring pair-tunneling term.

\par The single-particle inter-ring tunneling amplitude $t_\perp$ can, in principle, be obtained microscopically from the overlap of neighboring domain-wall wavefunctions,
\begin{equation}
t_\perp=\langle \Psi_i|H_{\rm TB}|\Psi_j\rangle,
\end{equation}
where $|\Psi_i\rangle$ and $|\Psi_j\rangle$ denote domain-wall states localized on the neighboring domain walls and $H_{\rm TB}$ is the underlying tight-binding Hamiltonian.

% ==============================================
\section{Estimation of effective Coulomb interactions}\label{sec:appx_c}
\par We estimate the effective intra- and inter-ring interactions under screened Coulomb potentials. Estimating the screened Coulomb interactions, which scale as $V_0\propto\ln(D/w)$ and $U_0\propto\ln(D/d)$ for the intra- and inter-ring channels \cite{PhysRevLett.102.256803}, respectively, we focus on the physically relevant regime $d\lesssim w$, where the neighboring domain-wall states exhibit sufficient wavefunction overlap to support inter-ring tunneling. Here, $D$ is the screening length, $w$ is the transverse localization length of the domain-wall states, and $d$ is the separation between the domain walls. This yields $U_0>V_0$ and $\tilde{K}_{-}>1$ over a broad parameter range. As $d/w$ decreases, $U_0-V_0\propto\ln(w/d)$ increases. Nevertheless, the limit $d\ll w$ is not physically relevant here, as the neighboring domain-wall states would strongly hybridize and cease to form well-defined HLLs on individual domain walls. Correspondingly, requiring a finite $\tilde{K}_{-}$ restricts the physically accessible regime to $2(U_0-V_0)/\pi\hbar v_F<1$. Since $\left(2U_0/\pi\hbar v_F\right)/\left(1+2V_0/\pi\hbar v_F\right) < 1$ in this regime, we have
\begin{widetext}
\begin{align}
K^{-1}=\sqrt{1+\frac{2V_0}{\pi\hbar v_F}} \quad > \quad \frac{1}{2}\left(\tilde{K}_{+}^{-1}+\tilde{K}_{-}^{-1}\right) = \frac{1}{2}\left(\sqrt{1+\frac{2V_0}{\pi\hbar v_F} + \frac{2U_0}{\pi\hbar v_F}} + \sqrt{1+\frac{2V_0}{\pi\hbar v_F} - \frac{2U_0}{\pi\hbar v_F}}\right),
\end{align}
\end{widetext}
which restores Eq.~\eqref{eq:eta_hierarchy}.

\par To obtain a representative estimate of the interaction strength for the self-consistent computation, we consider a screened-Coulomb model in the presence of metallic gates \cite{PhysRevLett.102.256803},
\begin{equation}
V_0 = \frac{e^2}{2\pi\epsilon_0\epsilon_{\rm eff}} \ln\!\left(\frac{D_g}{w}\right),\qquad U_0 = \frac{e^2} {2\pi\epsilon_0\epsilon_{\rm eff}} \ln\!\left(\frac{D_g}{d}\right),
\end{equation}
where $D_g$ is the gate distance. Taking representative values $\epsilon_{\rm eff}\approx 5$ \cite{PhysRevB.110.125429,C6TC01913G} and $\hbar v_F\approx 0.5~{\rm eV}\cdot{\rm nm}$ \cite{bordoloi2026}, together with $e^2/(4\pi\epsilon_0)\approx 1.44~{\rm eV}\cdot{\rm nm}$, yields
\begin{equation}\label{eq:U_V_est_1}
\frac{V_0}{\hbar v_F} \approx 1.152\ln\!\left(\frac{D_g}{w}\right), \qquad \frac{U_0}{\hbar v_F} \approx 1.152\ln\!\left(\frac{D_g}{d}\right).
\end{equation}
For representative ratios $5\lesssim D_g/w,D_g/d\lesssim15$, we obtain
\begin{equation}\label{eq:U_V_est_2}
1.2\lesssim\frac{2V_0}{\pi\hbar v_F},\,\frac{2U_0}{\pi\hbar v_F}\lesssim2,
\end{equation}
which motivates the parameter range $2V_0/\pi\hbar v_F,\,2U_0/\pi\hbar v_F\in[1.1,2.1]$ considered in the numerical computations shown in Fig.~\ref{fig:int_scan}. Note that based on Eqs.~\eqref{eq:U_V_est_1} and \eqref{eq:U_V_est_2}, $2(U_0-V_0)/\pi\hbar v_F\approx1$ indicates that $d\approx0.256w$. This suggests strong hybridization between the domain-wall states, implying non-perturbative inter-ring tunnelings.

% ==============================================
\section{Effective mass parameter $m_J$ in the infinite-size theory}\label{sec:appx_d}
\par To quantify the effect of finite-size fluctuations, we estimate the infinite-size counterpart of $m_J$ based on the Hamiltonian in Eq.~\eqref{eq:inf_H}. We first obtain the gap scale  $\Delta_{\rm gap}$ from the standard RG analysis of the infinite-size theory and then relate it to the corresponding mass parameter $m_J^{(\rm pin)}$. This estimate can also be derived within the variational framework \cite{book:Giamarchi}, but the RG derivation provides a more direct connection to the infinite-size strong-coupling scale. When $\tilde{J}_{\rm SC}$ reaches the strong-coupling region and pins the $\vartheta_-^{(\infty)}(r)$ field at $\vartheta_-^{(\infty)}(r)=0$, we can expand the pair-tunneling term to the quadratic order near $\vartheta_-^{(\infty)}(r)=0$, yielding an effective mass term:
\begin{equation}
    {H_{\rm m} = \frac{m_J^{(\rm pin)}}{(2\pi a)^2} \int_0^{l_s\to\infty} dr [\vartheta_-^{(\infty)}(r)]^2.}
\end{equation}
This leads to the action:
\begin{widetext}
\begin{equation}
    S^{(\rm pin)} = \frac{\hbar \tilde{K}_{-}}{2\pi}\int dr \int d\tau \left[ \frac{1}{\tilde{u}_{-}}\left(\partial_\tau\vartheta_-^{(\infty)}(r,\tau)\right)^2 + \tilde{u}_{-}\left(\partial_r\vartheta_-^{(\infty)}(r,\tau)\right)^2 + {\frac{m_J^{(\rm pin)}}{2\pi\hbar\tilde K_- a^2}\left(\vartheta_-^{(\infty)}(r,\tau)\right)^2} \right].
\end{equation}
\end{widetext}
From the corresponding dispersion relation, we can estimate the spectral gap $\Delta_{\rm gap}$ by using
\begin{equation}\label{eq:gap_SM}
    {\Delta_{\rm gap} = \sqrt{\frac{\hbar\tilde{u}_{-}}{2\pi\tilde{K}_{-}a^2}m_J^{(\rm pin)}}.}
\end{equation}
On the other hand, when $\tilde{J}_{\rm SC}$ reaches the strong-coupling region through the RG flow, $\Delta_{\rm gap}$ is renormalized as \cite{book:Giamarchi}
\begin{equation}\label{eq:gap_renorm_SM}
    \Delta_{\rm gap} \simeq \frac{\hbar \tilde{u}_{-}}{a}[\tilde{J}_{\rm SC}(0)]^{1/\left[2(1-\tilde{K}_{-}^{-1})\right]}.
\end{equation}
Therefore, combining Eqs.~\eqref{eq:gap_SM} and \eqref{eq:gap_renorm_SM}, we get
\begin{equation}\label{eq:m_infty_SM}
    {m_J^{(\rm pin)} = 2\pi\hbar\tilde{u}_{-}\tilde{K}_{-}[\tilde{J}_{\rm SC}(0)]^{2/\left[2(1-\tilde{K}_{-}^{-1})\right]}.}
\end{equation}
\end{appendix}
% \nocite{*}

% ==============================================
\bibliography{apssamp}% Produces the bibliography via BibTeX.
% ==============================================
\setcounter{equation}{0}
\setcounter{figure}{0}
\setcounter{table}{0}

\renewcommand{\theequation}{S\arabic{equation}}
\renewcommand{\thefigure}{S\arabic{figure}}
\renewcommand{\thetable}{S\arabic{table}}
\renewcommand{\bibnumfmt}[1]{[S#1]}
\renewcommand{\citenumfont}[1]{S#1}
\newcommand{\bk}{\boldsymbol\kappa}

\newcommand{\beginsupplement}{%
  \setcounter{equation}{0}
  \renewcommand{\theequation}{S\arabic{equation}}%
  \setcounter{table}{0}
  \renewcommand{\thetable}{S\arabic{table}}%
  \setcounter{figure}{0}
  \renewcommand{\thefigure}{S\arabic{figure}}%
  \setcounter{section}{0}
  \renewcommand{\thesection}{S\Roman{section}}%
  \setcounter{subsection}{0}
  \renewcommand{\thesubsection}{S\Roman{section}.\Alph{subsection}}%
}

\clearpage
\pagebreak
\beginsupplement
\begin{widetext}
\begin{center}
\textbf{\large Supplemental Materials: Helical Domain-Wall-Ring Networks Reshape Superconducting Correlations}
\end{center}

\SMtableofcontents
\enableSMtoc
% ===================================================
\section{Full HLL Hamiltonian in momentum space}
\par We discuss details of mode expansion for the full HLL Hamiltonians, which include the Hamiltonian $H_{\rm HLL}$ and the inter-ring density-density interaction $H_{\rm inter}$. We begin with the bosonized Hamiltonian of the inter-ring density-density interaction:
\begin{align}\label{eq:SM_H_inter}
    H_{\rm inter} = & {\frac{U_0}{2\pi^2}\sum_{\mathbf{R}}\sum_{j=0}^{5}\int_{jl_s}^{(j+1)l_s}dr \bigg[\partial_r\varphi_{\mathbf{R}}(r)\left.\partial_s\varphi_{\mathbf{R}+\mathbf{a}_j}(s)\right|_{s=f_j(r)}+\partial_r\varphi_{\mathbf{R}-\mathbf{a}_j}(r)\left.\partial_s\varphi_{\mathbf{R}}(s)\right|_{s=f_j(r)}\bigg]} \notag
    \\ = & {-\frac{U_0}{2\pi^2}\sum_{\mathbf{R}}\sum_{j=0}^{5}\int_{jl_s}^{(j+1)l_s}dr\,\bigg[\partial_r\varphi_{\mathbf{R}}(r)\frac{d}{dr}\varphi_{\mathbf{R}+\mathbf{a}_j}(f_j(r)) +\partial_r\varphi_{\mathbf{R}-\mathbf{a}_j}(r)\frac{d}{dr}\varphi_{\mathbf{R}}(f_j(r))\bigg]}. 
\end{align}
{The derivative $\partial_s$ acts on the neighboring ring's intrinsic coordinate, where $d\varphi(f_j(r))/dr=-\partial_s\varphi(s)|_{s=f_j(r)}$ with $f_j(r)=(2j+4)l_s-r\pmod L$. The two displayed terms are related by a translation of the ring-center summation; this symmetrized form is equivalent to the directed inter-ring interaction of the main text.} As noted in the main text, the bosonic fields are decomposed as:
\begin{align}
    \varphi_{\mathbf{R}}(r) & = \frac{1}{\sqrt{NL}}\sum_{\mathbf{p}=(\mathbf{q},k_n)}e^{i\mathbf{p}\cdot\mathbf{\mathcal{R}}}\varphi_{\mathbf{p}},
    \\ \vartheta_{\mathbf{R}}(r) & = \frac{1}{\sqrt{NL}}\sum_{\mathbf{p}=(\mathbf{q},k_n)}e^{i\mathbf{p}\cdot\mathbf{\mathcal{R}}}\vartheta_{\mathbf{p}},
\end{align}
where zero modes are excluded. Substituting the mode expansion into Eq.~\eqref{eq:SM_H_inter} yields
\begin{align}
H_{\rm inter} = {-\frac{U_0}{2\pi^2N}\sum_{\mathbf{R}}\sum_{j=0}^{5}\sum_{\mathbf{p}}\sum_{\mathbf{p}'}k_n k_{n'}e^{i(\mathbf{q}+\mathbf{q}')\cdot\mathbf{R}}\left[e^{i\mathbf{q}'\cdot\mathbf{a}_j}e^{i(2j+4)l_s k_{n'}} I_j(n,n') + e^{-i\mathbf{q}'\cdot\mathbf{a}_j}e^{i(2j+4)l_s k_{n}} I_j(n',n)\right]\varphi_{\mathbf{p}}\varphi_{\mathbf{p}'},}
\end{align}
where
\begin{align}
I_j(n,n') = \int_{jl_s}^{(j+1)l_s}\frac{dr}{L}\,e^{i(k_n-k_{n'})r}=e^{i(k_n-k_{n'})jl_s}\frac{e^{i(k_n-k_{n'})l_s}-1}{i{L}(k_n-k_{n'})} = e^{i\frac{\pi (n-n')(2j+1)}{6}}\frac{\sin(\frac{\pi (n-n')}{6})}{\pi (n-n')}, \label{eq:I_n_n}
\end{align}
which has the value $1/6$ for $n=n'$. Performing the summation over $\mathbf{R}$ yields
\begin{align}
\sum_{\mathbf{R}}e^{i(\mathbf{q}+\mathbf{q}')\cdot\mathbf{R}}=N\,\delta_{\mathbf{q}',-\mathbf{q}},
\end{align}
such that
\begin{align}
H_{\rm inter}={-\frac{U_0}{2\pi^2}\sum_{j=0}^{5}\sum_{\mathbf{q}}\sum_{n,n'}k_n k_{n'}\left[e^{-i\mathbf{q}\cdot\mathbf{a}_j}e^{i(2j+4)l_s k_{n'}} I_j(n,n') + e^{i\mathbf{q}\cdot\mathbf{a}_j}e^{i(2j+4)l_s k_{n}} I_j(n',n)\right]\varphi_{\mathbf{q},k_n}\varphi_{-\mathbf{q},k_{n'}},}
\end{align}
Since the interaction is present only in the finite overlap regions between the neighboring domain-wall rings, $I_j(n,n')$ is generally nonzero for $n\neq n'$, implying that different harmonics along the ring are coupled by the inter-ring interaction. The finite overlap regions thus generate off-diagonal matrix elements in $B_{\mathbf{q}}$, so the normal modes are hybridized combinations of different harmonics. Defining $k_m\equiv -k_{n'}$ yields
\begin{align}\label{eq:H_inter_derive_1}
H_{\rm inter}={\frac{U_0}{2\pi^2}\sum_{j=0}^{5}\sum_{\mathbf{q}}\sum_{n,m}k_n k_{m}\left[e^{-i\mathbf{q}\cdot\mathbf{a}_j}e^{-i(2j+4)l_s k_{m}} I_j(n,-m) + e^{i\mathbf{q}\cdot\mathbf{a}_j}e^{i(2j+4)l_s k_{n}} I_j(-m,n)\right]\varphi_{\mathbf{q},k_n}\varphi_{-\mathbf{q},-k_{m}},}
\end{align}
Since the bosonic fields are real, we have $\varphi_{-\mathbf{p}} = \varphi_{\mathbf{p}}^*$, Eq.~\eqref{eq:H_inter_derive_1} is equivalent to
\begin{align}
H_{\rm inter}={\frac{U_0}{2\pi^2}\sum_{j=0}^{5}\sum_{\mathbf{q}}\sum_{n,m}k_n k_{m}\left[e^{-i\mathbf{q}\cdot\mathbf{a}_j}e^{-i(2j+4)l_s k_{m}} \bar{I}_j(n,m) + e^{i\mathbf{q}\cdot\mathbf{a}_j}e^{i(2j+4)l_s k_{n}} \bar{I}_j^*(n,m)\right]\varphi_{\mathbf{q},k_n}\varphi_{\mathbf{q},k_{m}}^*,}
\end{align}
where 
\begin{equation}
\bar{I}_j(n,m) = e^{i(k_n+k_{m})jl_s}\frac{e^{i(k_n+k_{m})l_s}-1}{i{L}(k_n+k_{m})} = e^{i\frac{\pi (n+m)(2j+1)}{6}}\frac{\sin(\frac{\pi (n+m)}{6})}{\pi (n+m)}, \label{eq:I_n_m}
\end{equation}
which has the value $1/6$ for $n+m=0$. Introducing the vectors $\boldsymbol{\varphi}_{\mathbf{q}}=(\cdots,\varphi_{\mathbf{q},k_{-1}},\varphi_{\mathbf{q},k_1},\cdots)^T$ and $\boldsymbol{\vartheta}_{\mathbf{q}}=(\cdots,\vartheta_{\mathbf{q},k_{-1}},\vartheta_{\mathbf{q},k_1},\cdots)^T$, the full Hamiltonian becomes
\begin{align}\label{eq:H_HLL_mode_SM}
H_{\rm full\,HLL}=\frac{\hbar u}{2\pi}\sum_{\mathbf{q}}\left[\boldsymbol{\vartheta}_{\mathbf{q}}^{\dagger}A_{\mathbf{q}}\boldsymbol{\vartheta}_{\mathbf{q}}+\boldsymbol{\varphi}_{\mathbf{q}}^{\dagger}B_{\mathbf{q}}\boldsymbol{\varphi}_{\mathbf{q}}\right].
\end{align}
For the present density-density interaction, $A_{\mathbf{q}}$ remains diagonal in the harmonic index,
\begin{align}\label{eq:A_def}
(A_{\mathbf{q}})_{nm}=K(k_n)^2\delta_{nm},
\end{align}
whereas $B_{\mathbf{q}}$ contains both the intra-ring contribution and the inter-ring coupling:
\begin{align}
(B_{\mathbf{q}})_{nm} & = {\frac{1}{K}(k_n)^2\delta_{nm} + \frac{U_0}{\pi\hbar u}k_nk_m\frac{\sin(\frac{\pi(n+m)}{6})}{\pi(n+m)}\sum_{j=0}^{5}\bigg[e^{-i\mathbf{q}\cdot\mathbf{a}_j}e^{-i(2j+4)l_s k_n}e^{i\frac{\pi(n+m)(2j+1)}{6}}+e^{i\mathbf{q}\cdot\mathbf{a}_j}e^{i(2j+4)l_s k_m}e^{-i\frac{\pi(n+m)(2j+1)}{6}}\bigg]} \notag
\\ & = {\frac{1}{K}(k_n)^2\delta_{nm} + \frac{2U_0}{\pi\hbar u}k_nk_m\frac{\sin(\frac{\pi(n+m)}{6})}{\pi(n+m)}\sum_{j=0}^{5}e^{-i\frac{\pi}{6}(2j+4)(n-m)}\cos\!\left(\mathbf{q}\cdot\mathbf{a}_j+\frac{\pi}{2}(n+m)\right)}. \label{eq:B_def}
\end{align}

% ===================================================
\section{Self-consistent equation from the variational approach}
\par We derive the self-consistent equation for the mass parameter $m_J$ using a variational method. The results are summarized in Eqs.~\eqref{eq:m_self_SM} to \eqref{eq:mathcal_Y_mj}.

% -------------------------------------------------------------------------------
\subsection{Full HLL action}
\par First, we derive the Green's function for the full HLL by constructing its Lagrangian $L_{\rm full\,HLL}$ and action $S_{\rm full\,HLL}$. Starting from the canonical commutation relation
\begin{equation}
[\varphi_{\mathbf{R}}(r),\vartheta_{\mathbf{R}'}(r')] = -i\frac{\pi}{2}\mathrm{sgn}(r-r')\delta_{\mathbf{R},\mathbf{R}'},
\end{equation}
we differentiate with respect to $r$ and obtain
\begin{equation}
[\vartheta_{\mathbf{R}}(r),\partial_{r'}\varphi_{\mathbf{R}'}(r')] = i\pi\delta(r-r')\delta_{\mathbf{R},\mathbf{R}'}.
\end{equation}
Therefore, the canonical momentum conjugate to $\vartheta_{\mathbf{R}}(r)$ is
\begin{equation}
\Pi_{\vartheta,\mathbf{R}}(r) = \frac{1}{\pi}\partial_r\varphi_{\mathbf{R}}(r).
\end{equation}
In momentum space,
\begin{equation}\label{eq:Pi_q_def}
\Pi_{\vartheta,\mathbf{q}} = {\frac{i}{\pi}\mathcal{K}\boldsymbol{\varphi}_{\mathbf{q}}},
\end{equation}
where
\begin{equation}\label{eq:mathcal_K}
(\mathcal{K})_{nn'} = k_n\delta_{nn'}.
\end{equation}
Hence,
\begin{equation}
\boldsymbol{\varphi}_{\mathbf{q}} = {-i\pi\mathcal{K}^{-1}\boldsymbol{\Pi}_{\vartheta,\mathbf{q}}}.
\end{equation}
Substituting this into Eq.~\eqref{eq:H_HLL_mode_SM} gives
\begin{equation}
H_{\rm full\,HLL} = \frac{\hbar u}{2\pi}\sum_{\mathbf{q}} \left[ \boldsymbol{\Pi}_{\vartheta,\mathbf{q}}^{\dagger}D_{\mathbf{q}}\boldsymbol{\Pi}_{\vartheta,\mathbf{q}} + \boldsymbol{\vartheta}_{\mathbf{q}}^{\dagger}A_{\mathbf{q}}\boldsymbol{\vartheta}_{\mathbf{q}} \right],
\end{equation}
where
\begin{equation}\label{eq:D_q_def}
D_{\mathbf{q}} = \pi^2\mathcal{K}^{-1}B_{\mathbf{q}}\mathcal{K}^{-1}.
\end{equation}
The corresponding Lagrangian is
\begin{equation}
L_{\rm full\,HLL} = \sum_{\mathbf{q}}\boldsymbol{\Pi}_{\vartheta,\mathbf{q}}^{\dagger}\dot{\boldsymbol{\vartheta}}_{\mathbf{q}} - H_{\rm full\,HLL}.
\end{equation}
Variation with respect to $\boldsymbol{\Pi}_{\vartheta,\mathbf{q}}^{\dagger}$ yields
\begin{equation}
\dot{\boldsymbol{\vartheta}}_{\mathbf{q}} = \frac{\hbar u}{\pi}D_{\mathbf{q}}\boldsymbol{\Pi}_{\vartheta,\mathbf{q}},
\end{equation}
or equivalently
\begin{equation}
\boldsymbol{\Pi}_{\vartheta,\mathbf{q}} = \frac{\pi}{\hbar u}D_{\mathbf{q}}^{-1}\dot{\boldsymbol{\vartheta}}_{\mathbf{q}}.
\end{equation}
Substituting back into the Lagrangian gives
\begin{equation}
L_{\rm full\,HLL} = \frac{\pi}{2\hbar u} \sum_{\mathbf{q}}\dot{\boldsymbol{\vartheta}}_{\mathbf{q}}^{\dagger}D_{\mathbf{q}}^{-1}\dot{\boldsymbol{\vartheta}}_{\mathbf{q}} - \frac{\hbar u}{2\pi}\sum_{\mathbf{q}}\boldsymbol{\vartheta}_{\mathbf{q}}^{\dagger}A_{\mathbf{q}}\boldsymbol{\vartheta}_{\mathbf{q}}.
\end{equation}
After a Wick rotation $t\rightarrow -i\tau$, the Euclidean action becomes
\begin{equation}
S_{\rm full\,HLL} = \int d\tau\sum_{\mathbf{q}} \left[ \frac{\pi}{2\hbar u}\partial_{\tau}\boldsymbol{\vartheta}_{\mathbf{q}}^{\dagger}D_{\mathbf{q}}^{-1}\partial_{\tau}\boldsymbol{\vartheta}_{\mathbf{q}} + \frac{\hbar u}{2\pi}\boldsymbol{\vartheta}_{\mathbf{q}}^{\dagger}A_{\mathbf{q}}\boldsymbol{\vartheta}_{\mathbf{q}} \right].
\end{equation}
Introducing the Matsubara expansion
\begin{equation}
\boldsymbol{\vartheta}_{\mathbf{q}}(\tau) = \frac{1}{\sqrt{\beta}}\sum_{\omega_m}e^{-i\omega_m\tau}\boldsymbol{\vartheta}_{\mathbf{q}}(\omega_m),
\end{equation}
where $\omega_m=2\pi m/\beta$ is the bosonic Matsubara frequency with $m\in\mathbb{Z}$, we obtain
\begin{align}
S_{\rm full\,HLL} = \frac{1}{2}\sum_{\omega_m,\mathbf{q}}\boldsymbol{\vartheta}_{\mathbf{q}}^{\dagger}(\omega_m) \left[ \frac{\pi}{\hbar u}\omega_m^2D_{\mathbf{q}}^{-1} + \frac{\hbar u}{\pi}A_{\mathbf{q}} \right] \boldsymbol{\vartheta}_{\mathbf{q}}(\omega_m) \equiv \frac{1}{2}\sum_{\omega_m,\mathbf{q}}\boldsymbol{\vartheta}_{\mathbf{q}}^{\dagger}(\omega_m)\mathcal{G}_{\rm full\,HLL,\mathbf{q}}^{-1}(\omega_m)\boldsymbol{\vartheta}_{\mathbf{q}}(\omega_m),
\end{align}
where
\begin{equation}\label{eq:G_HLL_def}
\mathcal{G}_{\rm full\,HLL,\mathbf{q}}^{-1}(\omega_m) = \frac{\pi}{\hbar u}\omega_m^2D_{\mathbf{q}}^{-1} + \frac{\hbar u}{\pi}A_{\mathbf{q}} = \frac{1}{\pi\hbar u}\omega_m^2\mathcal{K} B_{\mathbf{q}}^{-1}\mathcal{K} + \frac{\hbar u}{\pi}A_{\mathbf{q}}. 
\end{equation}
% -------------------------------------------------------------------------------
\subsection{Pair-tunneling action, trial action, and the variational free energy}
\par We write the total action as $S=S_{\rm full\,HLL}+S_T$, where
\begin{align}
S_T & = -J_0\sum_{\mathbf{R}}\sum_{j=0}^{5}\int d\tau\int_{jl_s}^{(j+1)l_s}dr\,\cos\left[2\Theta_{\mathbf{R}+\mathbf{a}_j}(r,\tau)\right],
\\ \Theta_{\mathbf{R}+\mathbf{a}_j}(r) & \equiv {\vartheta_{\mathbf{R}+\mathbf{a}_j}(f_j(r))-\vartheta_{\mathbf{R}}(r)}. \label{eq:Theta_def}
\end{align}
with $J_0=J_{\rm SC}/(2\pi a)^2$. Upon including imaginary time, the phase-difference field becomes
\begin{equation}
\Theta_{\mathbf{R}+\mathbf{a}_j}(r,\tau)=\frac{1}{\sqrt{\beta NL}}\sum_{\omega_m,\mathbf{q},n}e^{i\mathbf{q}\cdot\mathbf{R}}e^{-i\omega_m\tau}\Lambda_{j,n}(\mathbf{q};r)\vartheta_{\mathbf{q},k_n}(\omega_m),
\end{equation}
where
\begin{equation}\label{eq:Lambda_def}
{\Lambda_{j,n}(\mathbf{q};r)=e^{i\mathbf{q}\cdot\mathbf{a}_j}e^{ik_n f_j(r)}-e^{ik_n r}.}
\end{equation}
To proceed with the variational method, we choose a Gaussian trial action
\begin{equation}
S_0=\frac{1}{2}\sum_{\omega_m,\mathbf{q}}\boldsymbol{\vartheta}_{\mathbf{q}}^{\dagger}(\omega_m)\mathcal{G}_{\mathbf{q}}^{-1}(\omega_m)\boldsymbol{\vartheta}_{\mathbf{q}}(\omega_m), 
\end{equation}
where
\begin{equation}\label{eq:G_def}
[\mathcal{G}_{\mathbf{q}}(\omega_m)]_{nn'}=\left\langle\vartheta_{\mathbf{q},k_n}(\omega_m)\vartheta_{-\mathbf{q},-k_{n'}}(-\omega_m)\right\rangle_0
\end{equation}
is the variational Green's function. Therefore, the Gaussian contribution to the variational free energy $F_{\rm var} = F_0 + T\langle S-S_0\rangle_0$ is
\begin{equation}
F_{\rm Gauss, \rm var}[G]=-\frac{T}{2}\sum_{\omega_m,\mathbf{q}}{\rm Tr}\ln \mathcal{G}_{\mathbf{q}}+\frac{T}{2}\sum_{\omega_m,\mathbf{q}}{\rm Tr}\left[\mathcal{G}_{\rm full\,HLL,\mathbf{q}}^{-1}\mathcal{G}_{\mathbf{q}}\right],
\end{equation}
up to an irrelevant constant. Since the trial action is Gaussian,
\begin{equation}
\left\langle\cos\left[2\Theta_{\mathbf{R}+\mathbf{a}_j}(r,\tau)\right]\right\rangle_0=\exp\left[-2C_{\Theta}^{(j)}(r)\right],
\end{equation}
where
\begin{equation}\label{eq:C_phi_SM}
C_{\Theta}^{(j)}(r)=\left\langle\Theta_{\mathbf{R}+\mathbf{a}_j}(r,\tau)^2\right\rangle_0=\frac{1}{\beta NL}\sum_{\omega_m,\mathbf{q}}\sum_{n,n'}\Lambda_{j,n}(\mathbf{q};r)[\mathcal{G}_{\mathbf{q}}(\omega_m)]_{nn'}\Lambda_{j,-n'}(-\mathbf{q};r).
\end{equation}
Therefore,
\begin{equation}
\langle S_T\rangle_0=-J_0\sum_{\mathbf{R}}\sum_{j=0}^{5}\int d\tau\int_{jl_s}^{(j+1)l_s}dr\,e^{-2C_{\Theta}^{(j)}(r)}.
\end{equation}
Thus, the total variational free energy is
\begin{equation}
F_{\rm var}[G]=-\frac{T}{2}\sum_{\omega_m,\mathbf{q}}{\rm Tr}\ln \mathcal{G}_{\mathbf{q}}+\frac{T}{2}\sum_{\omega_m,\mathbf{q}}{\rm Tr}\left[\mathcal{G}_{\rm full\,HLL,\mathbf{q}}^{-1}\mathcal{G}_{\mathbf{q}}\right]-TJ_0\sum_{\mathbf{R}}\sum_{j=0}^{5}\int d\tau\int_{jl_s}^{(j+1)l_s}dr\,e^{-2C_{\Theta}^{(j)}(r)}.
\end{equation}
% -------------------------------------------------------------------------------
\subsection{Self-consistency equation}
We now minimize $F_{\rm var}$ with respect to $[\mathcal{G}_{\mathbf{q}}(\omega_m)]_{nn'}$. The trace terms are varied using
\begin{equation}
\delta\,{\rm Tr}\ln \mathcal{G}_{\mathbf{q}}={\rm Tr}\left(\mathcal{G}_{\mathbf{q}}^{-1}\delta \mathcal{G}_{\mathbf{q}}\right),
\end{equation}
and
\begin{equation}
\delta\,{\rm Tr}\left[\mathcal{G}_{\rm full\,HLL,\mathbf{q}}^{-1}\mathcal{G}_{\mathbf{q}}\right]={\rm Tr}\left[\mathcal{G}_{\rm full\,HLL,\mathbf{q}}^{-1}\delta \mathcal{G}_{\mathbf{q}}\right].
\end{equation}
For the cosine contribution, we use
\begin{equation}
\frac{\delta C_{\Theta}^{(j)}(r)}{\delta [\mathcal{G}_{\mathbf{q}}(\omega_m)]_{nn'}}=\frac{1}{\beta NL}\Lambda_{j,n}(\mathbf{q};r)\Lambda_{j,-n'}(-\mathbf{q};r).
\end{equation}
Therefore,
\begin{equation}
\frac{\delta}{\delta [\mathcal{G}_{\mathbf{q}}(\omega_m)]_{nn'}}\left[-TJ_0\sum_{\mathbf{R},j}\int d\tau\int_{jl_s}^{(j+1)l_s}dr\,e^{-2C_{\Theta}^{(j)}(r)}\right]=\frac{2TJ_0}{\beta NL}\sum_{\mathbf{R},j}\int d\tau\int_{jl_s}^{(j+1)l_s}dr\,e^{-2C_{\Theta}^{(j)}(r)}\Lambda_{j,n}(\mathbf{q};r)\Lambda_{j,-n'}(-\mathbf{q};r).
\end{equation}
Using $\sum_{\mathbf{R}}=N$ and $\int d\tau=\beta$, the stationarity condition $\delta F_{\rm var}/\delta [\mathcal{G}_{\mathbf{q}}(\omega_m)]_{nn'}=0$ yields
\begin{equation}
-\frac{T}{2}[\mathcal{G}_{\mathbf{q}}^{-1}(\omega_m)]_{n'n}+\frac{T}{2}\mathcal [G_{\rm HLL,\mathbf{q}}^{-1}(\omega_m)]_{n'n}+\frac{2TJ_0}{L}\sum_{j=0}^{5}\int_{jl_s}^{(j+1)l_s}dr\,e^{-2C_{\Theta}^{(j)}(r)}\Lambda_{j,n}(\mathbf{q};r)\Lambda_{j,-n'}(-\mathbf{q};r)=0.
\end{equation}
Equivalently,
\begin{equation}\label{eq:G_self_SM}
[\mathcal{G}_{\mathbf{q}}^{-1}(\omega_m)]_{nn'}=[\mathcal{G}_{\rm full\,HLL,\mathbf{q}}^{-1}(\omega_m)]_{nn'}+J_0(\tilde{\mathcal{M}}_{\mathbf{q}})_{nn'},
\end{equation}
with
\begin{equation}
(\tilde{\mathcal{M}}_{\mathbf{q}})_{nn'}=\frac{4}{L}\sum_{j=0}^{5}\int_{jl_s}^{(j+1)l_s}dr\,e^{-2C_{\Theta}^{(j)}(r)}\Lambda_{j,n'}(\mathbf{q};r)\Lambda_{j,-n}(-\mathbf{q};r).
\end{equation}
Since Eq.~\eqref{eq:G_self_SM} generates a frequency-independent self-energy, the variational Green's function can be written as
\begin{equation}
\mathcal{G}_{\mathbf{q}}^{-1}(\omega_m) = \omega_m^2\mathcal{X}_{\mathbf{q}} + \mathcal{Y}_{\mathbf{q}}.
\end{equation}
Therefore, Eq.~\eqref{eq:C_phi_SM} reduce to
\begin{align}
C_{\Theta}^{(j)}(r) & = \frac{1}{\beta NL}\sum_{\omega_m,\mathbf{q}}\sum_{n,n'}\Lambda_{j,n}(\mathbf{q};r) \left( \omega_m^2[\mathcal{X}_{\mathbf{q}}]_{nn'} + [\mathcal{Y}_{\mathbf{q}}]_{nn'} \right)^{-1} \Lambda_{j,-n'}(-\mathbf{q};r). \label{eq:C_phi_SM_2}
\end{align}
In the $T\to0$ limit, $\frac{1}{\beta}\sum_m\to\int \frac{d\omega}{2\pi}$, Eq.~\eqref{eq:C_phi_SM_2} becomes
\begin{equation}\label{eq:C_phi_T0}
C_{\Theta}^{(j)}(r)=\frac{1}{NL}\sum_{\mathbf{q}}\sum_{n,n'}\Lambda_{j,n}(\mathbf{q};r)\left[\int\frac{d\omega}{2\pi}\left(\omega^2\mathcal{X}_{\mathbf{q}}+\mathcal{Y}_{\mathbf{q}}\right)^{-1}\right]_{nn'}\Lambda_{j,-n'}(-\mathbf{q};r).
\end{equation}
To evaluate the frequency integral, we define
\begin{equation}
\Omega_{\mathbf{q}}^2=\mathcal{X}_{\mathbf{q}}^{-1/2}\mathcal{Y}_{\mathbf{q}}\mathcal{X}_{\mathbf{q}}^{-1/2}.
\end{equation}
Then
\begin{equation}
\omega^2\mathcal{X}_{\mathbf{q}}+\mathcal{Y}_{\mathbf{q}}=\mathcal{X}_{\mathbf{q}}^{1/2}\left(\omega^2+\Omega_{\mathbf{q}}^2\right)\mathcal{X}_{\mathbf{q}}^{1/2},
\end{equation}
and hence
\begin{equation}
\left(\omega^2\mathcal{X}_{\mathbf{q}}+\mathcal{Y}_{\mathbf{q}}\right)^{-1}=\mathcal{X}_{\mathbf{q}}^{-1/2}\left(\omega^2+\Omega_{\mathbf{q}}^2\right)^{-1}\mathcal{X}_{\mathbf{q}}^{-1/2}.
\end{equation}
Using
\begin{equation}
\int\frac{d\omega}{2\pi}\frac{1}{\omega^2+\Omega_{\mathbf{q}}^2}=\frac{1}{2\Omega_{\mathbf{q}}},
\end{equation}
we obtain
\begin{equation}
\int\frac{d\omega}{2\pi}\left(\omega^2\mathcal{X}_{\mathbf{q}}+\mathcal{Y}_{\mathbf{q}}\right)^{-1}=\frac{1}{2}\mathcal{X}_{\mathbf{q}}^{-1/2}\left(\mathcal{X}_{\mathbf{q}}^{-1/2}\mathcal{Y}_{\mathbf{q}}\mathcal{X}_{\mathbf{q}}^{-1/2}\right)^{-1/2}\mathcal{X}_{\mathbf{q}}^{-1/2}.
\end{equation}
Therefore,
\begin{equation}\label{eq:C_phi_G}
C_{\Theta}^{(j)}(r)=\frac{1}{NL}\sum_{\mathbf{q}}\sum_{n,n'}\Lambda_{j,n}(\mathbf{q};r)[\mathcal{G}_{\mathbf{q}}]_{nn'}\Lambda_{j,-n'}(-\mathbf{q};r),
\end{equation}
where
\begin{equation}\label{eq:G_matrix}
\mathcal{G}_{\mathbf{q}}=\frac{1}{2}\mathcal{X}_{\mathbf{q}}^{-1/2}\left(\mathcal{X}_{\mathbf{q}}^{-1/2}\mathcal{Y}_{\mathbf{q}}\mathcal{X}_{\mathbf{q}}^{-1/2}\right)^{-1/2}\mathcal{X}_{\mathbf{q}}^{-1/2}.
\end{equation}
Combining Eqs.~\eqref{eq:C_phi_G} and \eqref{eq:G_matrix}, Eq.~\eqref{eq:G_self_SM} reduces to the frequency independent form
\begin{align}
[\mathcal{X}_{\mathbf{q}}]_{nn'} & =\frac{1}{\pi\hbar u}[\mathcal{K} B_{\mathbf{q}}^{-1}\mathcal{K}]_{nn'}, \label{eq:mathcal_X}
\\ [\mathcal{Y}_{\mathbf{q}}]_{nn'} & =\frac{\hbar u}{\pi}[A_{\mathbf{q}}]_{nn'}+\frac{4J_0}{L}\sum_{j=0}^{5}\int_{jl_s}^{(j+1)l_s}dr\,e^{-2C_{\Theta}^{(j)}(r)}\Lambda_{j,n'}(\mathbf{q};r)\Lambda_{j,-n}(-\mathbf{q};r),
\\ C_{\Theta}^{(j)}(r) & =\frac{1}{2NL}\sum_{\mathbf{q}}\sum_{n,n'}\Lambda_{j,n}(\mathbf{q};r)[\mathcal{X}_{\mathbf{q}}^{-1/2}\left(\mathcal{X}_{\mathbf{q}}^{-1/2}\mathcal{Y}_{\mathbf{q}}\mathcal{X}_{\mathbf{q}}^{-1/2}\right)^{-1/2}\mathcal{X}_{\mathbf{q}}^{-1/2}]_{nn'}\Lambda_{j,-n'}(-\mathbf{q};r),
\end{align}
where $A_{\mathbf{q}}$, $B_{\mathbf{q}}$, and $\mathcal{K}$ are defined in Eqs.~\eqref{eq:A_def}, \eqref{eq:B_def}, and \eqref{eq:mathcal_K}, respectively. Motivated by the quadratic expansion of the two-particle tunneling term around its energy minimum, we model $\mathcal{G}_{\mathbf{q}}^{-1}(\omega_m)$ as
\begin{equation}
[\mathcal{G}_{\mathbf{q}}^{-1}(\omega_m)]_{nn'}=[\mathcal{G}_{\rm full\,HLL,\mathbf{q}}^{-1}(\omega_m)]_{nn'}+\frac{m_J}{(2\pi a)^2}(\mathcal{M}_{\mathbf{q}})_{nn'},
\end{equation}
where
\begin{align}
    (\mathcal{M}_{\mathbf{q}})_{nn'} & = {\sum_{j=0}^{5}\int_{jl_s}^{(j+1)l_s}\frac{dr}{L}\Lambda_{j,n'}(\mathbf{q};r)\Lambda_{j,-n}(-\mathbf{q};r)}, \notag
    \\ & = {2\delta_{nn'} - 2\frac{\sin\!\left[\frac{\pi}{6}(n+n')\right]}{\pi(n+n')}\sum_{j=0}^{5}e^{-i\frac{\pi}{6}(2j+4)(n-n')}\cos\!\left[\mathbf{q}\cdot\mathbf{a}_j+\frac{\pi}{2}(n+n')\right]}.
\end{align}
This yields a self-consistent equation for the variational mass parameter $m_J$:
\begin{align}
    m_J & = J_{\rm SC}\frac{\sum_{\mathbf{q}}\text{Tr}[\mathcal{M}_{\mathbf{q}}^\dagger\tilde{\mathcal{M}}_{\mathbf{q}}(m_J)]}{\sum_{\mathbf{q}}\text{Tr}[\mathcal{M}_{\mathbf{q}}^\dagger\mathcal{M}_{\mathbf{q}}]}, \label{eq:m_self_SM}
\end{align}
where
\begin{align}
    [\tilde{\mathcal{M}}_{\mathbf{q}}(m_J)]_{nn'} & = \frac{4}{L}\sum_{j=0}^{5}\int_{jl_s}^{(j+1)l_s}dr\,e^{-2C_{\Theta}^{(j)}(r; m_J)}\Lambda_{j,n'}(\mathbf{q};r)\Lambda_{j,-n}(-\mathbf{q};r), \label{eq:M_var_SM}
    \\ C_{\Theta}^{(j)}(r; m_J) & =\frac{1}{2NL}\sum_{\mathbf{q}}\sum_{n,n'}\Lambda_{j,n}(\mathbf{q};r)[\mathcal{X}_{\mathbf{q}}^{-1/2}\left(\mathcal{X}_{\mathbf{q}}^{-1/2}\mathcal{Y}_{\mathbf{q}}(m_J)\mathcal{X}_{\mathbf{q}}^{-1/2}\right)^{-1/2}\mathcal{X}_{\mathbf{q}}^{-1/2}]_{nn'}\Lambda_{j,-n'}(-\mathbf{q};r),
    \\ [\mathcal{Y}_{\mathbf{q}}(m_J)]_{nn'} & =\frac{\hbar u}{\pi}[A_{\mathbf{q}}]_{nn'}+\frac{m_J}{(2\pi a)^2}(\mathcal{M}_{\mathbf{q}})_{nn'}. \label{eq:mathcal_Y_mj}
\end{align}
Here, $A_{\mathbf{q}}$, $\Lambda_{j,n}(\mathbf{q};r)$, and $\mathcal{X}_{\mathbf{q}}$ are defined in Eqs.~\eqref{eq:A_def}, \eqref{eq:Lambda_def}, and \eqref{eq:mathcal_X}, respectively.

\par Unless otherwise specified, all results were obtained using a $31\times31$ $\mathbf{q}$-point mesh and 40 integration points along the domain-wall coordinate $r$. The self-consistent procedure was iterated until a convergence criterion of $10^{-10}$ was reached. Further increasing the mesh densities or tightening the convergence criterion yielded no significant changes. 

% ===================================================
\section{SC correlation function}
\par We discuss how to compute the SC correlation function and estimate the SC scaling dimension $\eta_{\rm SC}$ in the domain-wall-ring network. The SC correlation function is defined as
\begin{equation}
C_{\rm SC}(r)=\langle \mathcal{O}_{\rm SC}^{\dagger}(r)\mathcal{O}_{\rm SC}(0)\rangle_0,
\end{equation}
where the operator characterizing the SC instability is
\begin{equation}
\mathcal{O}_{\rm SC}(r)=\mathcal{U}_{\mathbf{R},+}(r)\mathcal{U}_{\mathbf{R},-}(r)-\mathcal{U}_{\mathbf{R},-}(r)\mathcal{U}_{\mathbf{R},+}(r)\sim e^{i2\vartheta_{\mathbf{R}}(r)}.
\end{equation}
Using bosonization,
\begin{equation}
C_{\rm SC}(r)\sim\left\langle e^{-i2\vartheta_{\mathbf{R}}(r)}e^{i2\vartheta_{\mathbf{R}}(0)}\right\rangle_0 = e^{-2\langle [\vartheta_{\mathbf{R}}(r)-\vartheta_{\mathbf{R}}(0)]^2\rangle_0},
\end{equation}
where the second equality is obtained by evaluating the expectation value with the Gaussian trial action $S_0$. Motivated by the finite-size scaling form of correlation functions in a periodic geometry \cite{Cardy_1984}, we fit
\begin{equation}
\langle [\vartheta_{\mathbf{R}}(r)-\vartheta_{\mathbf{R}}(0)]^2\rangle_0 =\eta_{\rm SC}\ln\left[\frac{L}{\pi a}\sin\left(\frac{\pi r}{L}\right)\right]+\zeta,
\end{equation}
where $\zeta$ is a non-universal constant.

\par To compute $\langle [\vartheta_{\mathbf{R}}(r)-\vartheta_{\mathbf{R}}(0)]^2\rangle_0$, we mode expand the bosonic fields, yielding:
\begin{equation}
\vartheta_{\mathbf{R}}(r)-\vartheta_{\mathbf{R}}(0)=\frac{1}{\sqrt{NL}}\sum_{\mathbf{q}}\sum_{n}e^{i\mathbf{q}\cdot \mathbf{R}}\left(e^{ik_n r}-1\right)\vartheta_{\mathbf{q},k_n}.
\end{equation}
The phase fluctuation is then given by
\begin{equation}
\langle [\vartheta_{\mathbf{R}}(r)-\vartheta_{\mathbf{R}}(0)]^2\rangle_0=\frac{1}{NL}\sum_{\mathbf{q}}\sum_{n,n'}\left(e^{ik_n r}-1\right)\left(e^{-ik_{n'} r}-1\right)\langle \vartheta_{\mathbf{q},k_n}\vartheta_{-\mathbf{q},-k_{n'}}\rangle_0.
\end{equation}
Using Eqs.~\eqref{eq:G_def} and \eqref{eq:G_matrix} yields
\begin{equation}
\langle [\vartheta_{\mathbf{R}}(r)-\vartheta_{\mathbf{R}}(0)]^2\rangle_0=\frac{1}{2NL}\sum_{\mathbf{q}}\boldsymbol{\Gamma}^{\dagger}(r)\mathcal{X}_{\mathbf{q}}^{-1/2}\left(\mathcal{X}_{\mathbf{q}}^{-1/2}\mathcal{Y}_{\mathbf{q}}\mathcal{X}_{\mathbf{q}}^{-1/2}\right)^{-1/2}\mathcal{X}_{\mathbf{q}}^{-1/2}\boldsymbol{\Gamma}(r),
\end{equation}
where $\mathcal{X}_{\mathbf{q}}$ and $\mathcal{Y}_{\mathbf{q}}$ are defined in Eqs.~\eqref{eq:mathcal_X} and \eqref{eq:mathcal_Y_mj}, respectively, and
\begin{equation}
[\boldsymbol{\Gamma}(r)]_n=e^{-ik_n r}-1.
\end{equation}

% ===================================================
\section{Actions with zero modes}
\par We show that inclusion of zero modes in the mode expansion of the bosonic fields does not qualitatively alter the results obtained from the variational approach. We first write down the complete mode expansion of the bosonic fields:
\begin{align}
\varphi_{\mathbf{R}}(r) = & \frac{1}{\sqrt{NL}}\sum_{\mathbf{p}=(\mathbf{q},k_n\neq0)}e^{i\mathbf{p}\cdot\mathbf{\mathcal{R}}}\varphi_{\mathbf{p}}-\frac{\pi r}{L}\mathcal{N}_{\mathbf{R}} + \varphi_{0,\mathbf{R}},
\\ \vartheta_{\mathbf{R}}(r) = & \frac{1}{\sqrt{NL}}\sum_{\mathbf{p}=(\mathbf{q},k_n\neq0)}e^{i\mathbf{p}\cdot\mathbf{\mathcal{R}}}\vartheta_{\mathbf{p}}+\frac{\pi r}{L}\mathcal{J}_{\mathbf{R}} + \vartheta_{0,\mathbf{R}},
\end{align}
where $\mathbf{\mathcal{R}}=(\mathbf{R},r)$, $\mathcal{N}_{\mathbf{R}}=\mathcal{N}_{\mathbf{R}, +} + \mathcal{N}_{\mathbf{R}, -}$, and $\mathcal{J}_{\mathbf{R}} = \mathcal{N}_{\mathbf{R}, +} - \mathcal{N}_{\mathbf{R}, -}$. $\mathcal{N}_{\mathbf{R},+}$ and $\mathcal{N}_{\mathbf{R},-}$ denote the chiral charge operators for the clockwise- and counterclockwise-moving modes on the ring with its center at $\mathbf{R}$ \cite{LL_Haldane_1981}. The zero modes satisfy the commutation relation \cite{senechal1999}:
\begin{equation}
[\vartheta_{0,\mathbf{q}},\mathcal{N}_{\mathbf{q}'}]=i\delta_{\mathbf{q},-\mathbf{q}'},\qquad [\varphi_{0,\mathbf{q}},\mathcal{J}_{\mathbf{q}'}]=i\delta_{\mathbf{q},-\mathbf{q}'},
\end{equation}
where $\mathcal{N}_{\mathbf{q}}$ and $\mathcal{J}_{\mathbf{q}}$ denote the Fourier components of $\mathcal{N}_{\mathbf{R}}$ and $\mathcal{J}_{\mathbf{R}}$.

\par To construct the full HLL's action, we first derive its Hamiltonian. The zero-mode-only part of the HLL Hamiltonian gives
\begin{equation}
H_{\rm HLL}^{(\rm zm)}=\frac{\hbar u}{2\pi}\sum_{\mathbf{q}}\left[\frac{\pi^2}{KL}\mathcal{N}_{\mathbf{q}}\mathcal{N}_{-\mathbf{q}}+\frac{\pi^2K}{L}\mathcal{J}_{\mathbf{q}}\mathcal{J}_{-\mathbf{q}}\right].
\end{equation}
We next consider the inter-ring density-density interaction. {Using the intrinsic-coordinate charge density $\rho_{\mathbf R}(r)=-\partial_r\varphi_{\mathbf R}(r)/\pi$, $l_s=L/6$, and $\mathcal N_{\mathbf R}=N^{-1/2}\sum_{\mathbf q}e^{i\mathbf q\cdot\mathbf R}\mathcal N_{\mathbf q}$, the terms containing a zero mode are}
\begin{align}
H_{\rm inter}^{(\rm zm)} = &\ {-\frac{U_0}{\pi L\sqrt L}\sum_{\mathbf{q},n\ne0}\mathcal N_{-\mathbf q}\varphi_{\mathbf{q},k_n}\sum_{j=0}^{5}\bigg[e^{-i\mathbf q\cdot\mathbf a_j}\left(e^{ik_n(j+1)l_s}-e^{ik_njl_s}\right)-e^{i\mathbf q\cdot\mathbf a_j}e^{i(2j+4)l_s k_n}\left(e^{-ik_n(j+1)l_s}-e^{-ik_njl_s}\right)\bigg]}\notag
\\ &\ {+\frac{U_0}{6L}\sum_{\mathbf q}\sum_{j=0}^{5}\cos(\mathbf q\cdot\mathbf a_j)\mathcal N_{\mathbf q}\mathcal N_{-\mathbf q}}\notag
\\ \equiv &\ {\frac{U_0}{\pi L\sqrt L}\sum_{\mathbf{q},n\ne0}[\boldsymbol\lambda_{\mathbf q}]_n\mathcal N_{-\mathbf q}\varphi_{\mathbf{q},k_n}+\frac{U_0}{6L}\sum_{\mathbf q}\sum_{j=0}^{5}\cos(\mathbf q\cdot\mathbf a_j)\mathcal N_{\mathbf q}\mathcal N_{-\mathbf q}},\label{eq:H_inter_ZM}
\end{align}
where
\begin{equation}
{[\boldsymbol\lambda_{\mathbf q}]_n=-(e^{i\pi n/3}-1)\sum_{j=0}^{5}e^{i\pi nj/3}\left[e^{-i\mathbf q\cdot\mathbf a_j}+(-1)^n e^{i\mathbf q\cdot\mathbf a_j}\right]}. 
\end{equation}
{Define $\boldsymbol v_{\mathbf q}=\boldsymbol\lambda_{\mathbf q}^{*}$, with complex conjugation taken componentwise. Completing the square within a fixed charge sector by the shift $\boldsymbol\varphi_{\mathbf q}\mapsto\boldsymbol\varphi_{\mathbf q}-\frac{U_0}{\hbar uL\sqrt L}B_{\mathbf q}^{-1}\boldsymbol v_{\mathbf q}\mathcal N_{\mathbf q}$ gives}
\begin{align}
S_{\rm full\,HLL}=&\ {S_{\rm full\,HLL,oscillatory}^{\,\prime}+\int_0^\beta d\tau\,\frac{\pi\hbar u}{2L}\sum_{\mathbf q}\left[K_{\rm NN}^{\rm eff}(\mathbf q)\mathcal N_{\mathbf q}\mathcal N_{-\mathbf q}+K\mathcal J_{\mathbf q}\mathcal J_{-\mathbf q}\right]},\label{eq:S_zm_def}
\end{align}
{where}
\begin{align}
K_{\rm NN}^{\rm eff}(\mathbf q)&={\frac1K+\frac{U_0}{3\pi\hbar u}\sum_{j=0}^{5}\cos(\mathbf q\cdot\mathbf a_j)-\frac{U_0^2}{\pi^2(\hbar u)^2 L^2}\boldsymbol v_{\mathbf q}^{\dagger}B_{\mathbf q}^{-1}\boldsymbol v_{\mathbf q}}.
\end{align}
\par {To isolate the effect of oscillatory fluctuations on pair tunneling, we approximate $\vartheta_{0,\mathbf{R}+\mathbf{a}_j}\approx\vartheta_{0,\mathbf{R}}$. This configuration provides the largest pair-tunneling contribution for fixed oscillatory fluctuations, allowing us to quantify their suppressing effect. Within this approximation}, the complete action contains the terms $i\mathcal{N}_{\mathbf{q}}\partial_\tau\vartheta_{0,-\mathbf{q}}+i\mathcal{J}_{\mathbf{q}}\partial_\tau\varphi_{0,-\mathbf{q}}$. Integrating out $\vartheta_{0,-\mathbf{q}}$ and $\varphi_{0,-\mathbf{q}}$ constrains $\mathcal{N}_{\mathbf{q}}$ and $\mathcal{J}_{\mathbf{q}}$ to be time independent, after which the action reduces to Eq.~\eqref{eq:S_zm_def}. Since integrating out $\mathcal{N}_{\mathbf{q}}$ in Eq.~\eqref{eq:S_zm_def} does not change the Green's function of the oscillatory sector, quantities determined solely by {oscillatory} fluctuations, such as the SC scaling dimension $\eta_{\rm SC}$ and the phase-fluctuation correlation function $C_{\Theta}$, are unaffected by
$\mathcal{N}_{\mathbf{q}}$ {within this approximation}.

\par In contrast, $\mathcal{J}_{\mathbf{R}}$ enters the {phase-difference field} through
\begin{equation}
{
\begin{aligned}
\Theta_{\mathbf{R}+\mathbf{a}_j}(r) = &\,\Theta_{\mathbf{R}+\mathbf{a}_j}^{({\rm osc})}(r) + \frac{\pi}{L}\left[ \mathcal{J}_{\mathbf{R}+\mathbf{a}_j}f_j(r) -\mathcal{J}_{\mathbf{R}}r \right]\equiv\Theta_{\mathbf{R}+\mathbf{a}_j}^{({\rm osc})}(r)
+\Theta^{(\mathcal{J})}(r),
\end{aligned}
}
\end{equation}
{where $f_j(r)$ is evaluated modulo $L$.} Since the full partition function involves a sum over $\mathcal{J}_{\mathbf{R}}$, $Z=\sum_{\{\mathcal{J}_{\mathbf{R}}\}}
e^{-\beta F[\mathcal{J}_{\mathbf{R}}]}$, a complete determination of the thermodynamic ground state would require comparing the $\mathcal{J}_{\mathbf{R}}$-dependent variational free energies. In the present work, however, we are interested only in whether $\mathcal{J}_{\mathbf{R}}$ can enhance $m_J$. For each fixed $\{\mathcal{J}_{\mathbf{R}}\}$ sector, it contributes only a $c$-number phase shift to $\Theta_{\mathbf{R}+\mathbf{a}_j}(r)$, leading to $\left\langle \cos[2\Theta_{\mathbf{R}+\mathbf{a}_j}(r)] \right\rangle_0 =e^{-2C_\Theta^{(j)}(r)} \cos[2\Theta^{(\mathcal{J})}(r)]$ {within the adopted approximation}. Since $|\cos[2\Theta^{(\mathcal{J})}(r)]|\leq1$, a nonzero $\mathcal{J}_{\mathbf{R}}$ cannot produce a larger local pair-tunneling source than the $\mathcal{J}_{\mathbf{R}}=0$ sector. Thus, the $\mathcal{J}_{\mathbf{R}}=0$ sector gives the most favorable estimate for $m_J$. Including $\mathcal{J}_{\mathbf{R}}$ can only reduce the effective pair-tunneling
contribution {within this approximation} and, therefore, it does not weaken our conclusion that phase locking is strongly suppressed in the physically relevant interaction regime.

% ===================================================
\section{Other numerical results}\label{sec:num_SM}
Note that coefficients of determination ($R^2$) corresponding to the $\eta_{\rm SC}$-fit in Fig.~2(b) of the main text increase with decreasing $\theta_t$ (Fig.~\ref{fig:C_SC}(a)), indicating an increasingly accurate description by the scaling form at smaller twist angles. The $R^2$ values corresponding to the $\eta_{\rm SC}$-fit in Fig.~3(b) of the main text are shown in Fig.~\ref{fig:C_SC}(b).

\begin{figure}[h]
  \centering
  \centering
    \includegraphics[width=0.8\linewidth]{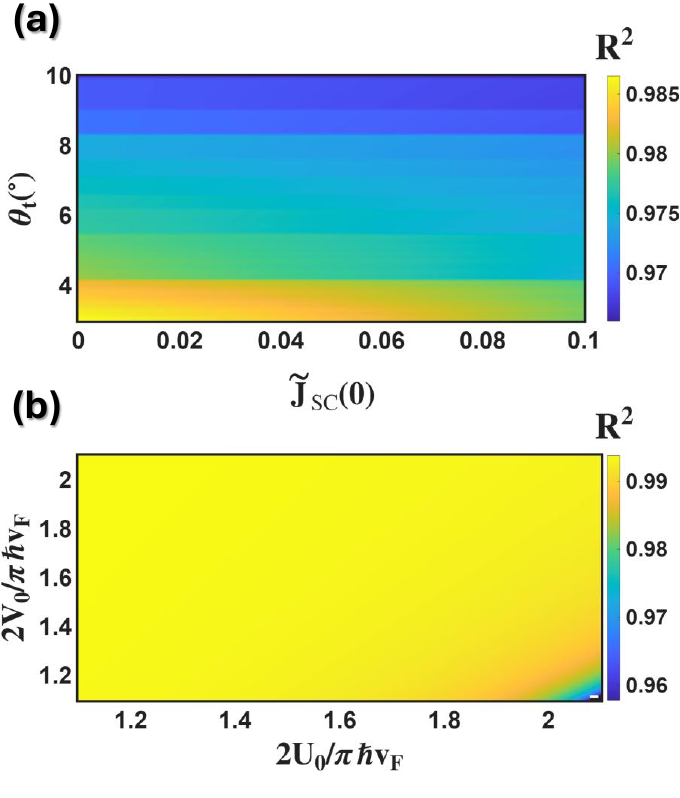}
  \caption{ (a,b) Coefficients of determination ($R^2$) associated with the fits shown in Figs.~2(b) and 4(b) of the main text, respectively. }
  \label{fig:C_SC}
\end{figure}

\end{widetext}
% ==============================================
\end{document}